\documentclass[aps,prl,twocolumn,showpacs,psfig,superscriptaddress,longbibliography]{revtex4-1}
\usepackage{textcomp}
\usepackage{times}
\usepackage{graphicx}
\usepackage{float}
\usepackage{latexsym,amsmath,amssymb,bm,euscript}
\usepackage{color}
\usepackage{subfigure}
\usepackage{epstopdf}
\usepackage[colorlinks=true,linkcolor=blue,citecolor=blue,urlcolor=blue]{hyperref}
\usepackage{hyperref}
\usepackage{soul}
\usepackage[normalem]{ulem}
\usepackage{mathrsfs}
\usepackage{amsmath,amssymb}
\usepackage{lettrine}
\usepackage{xfrac}
\usepackage{xspace}
\usepackage{textcomp}
\usepackage{wasysym}
\usepackage{xr}

\newcommand{\Fig}[1]{Fig.~\ref{#1}}

\graphicspath{{./}}

\AtBeginDocument{\newwrite\bibnotes
\def\bibnotesext{Notes.bib}
\immediate\openout\bibnotes=\jobname\bibnotesext
\immediate\write\bibnotes{@CONTROL{REVTEX41Control}}
\immediate\write\bibnotes{@CONTROL{apsrev41Control,author="08",editor="1",pages="1",title="0",year="1"}}
\if@filesw
\immediate\write\@auxout{\string\citation{apsrev41Control}}\fi
}

\begin{document}

\title{Exciton Proliferation and Fate of the Topological Mott Insulator in a
\\
Twisted Bilayer Graphene Lattice Model}
\author{Xiyue Lin}
\affiliation{CAS Key Laboratory of Theoretical Physics,
Institute of Theoretical Physics,
Chinese Academy of Sciences, Beijing 100190, China}
\affiliation{School of Physical Sciences, University of Chinese
Academy of Sciences, Beijng 100049, China}

\author{Bin-Bin Chen}
\email{bchenhku@hku.hk}
\affiliation{Department of Physics and HKU-UCAS Joint Institute of
Theoretical and Computational Physics, The University of Hong Kong,
Pokfulam Road, Hong Kong, China}
\affiliation{School of Physics, Beihang
University, Beijing 100191, China}

\author{Wei Li}
\affiliation{CAS Key Laboratory of Theoretical Physics,
Institute of Theoretical Physics,
Chinese Academy of Sciences, Beijing 100190, China}
\affiliation{School of
Physics, Beihang University, Beijing 100191, China}

\author{Zi Yang Meng}
\email{zymeng@hku.hk}
\affiliation{Department of Physics and HKU-UCAS Joint Institute of
Theoretical and Computational Physics, The University of Hong Kong,
Pokfulam Road, Hong Kong, China}

\author{Tao Shi}
\email{tshi@itp.ac.cn}
\affiliation{CAS Key Laboratory of Theoretical Physics,
Institute of Theoretical Physics,
Chinese Academy of Sciences, Beijing 100190, China}

\begin{abstract}
Topological Mott insulator (TMI) with spontaneous time-reversal
symmetry breaking and nonzero Chern number has been discovered 
in a real-space effective model for twisted bilayer graphene
(TBG) at 3/4 filling in the strong coupling limit~\cite{Chen2020TMI}.
However, the finite temperature properties of such a TMI state 
remain illusive. In this work, employing the state-of-the-art thermal 
tensor network and the perturbative field-theoretical approaches,
we obtain the finite-$T$ phase diagram and the dynamical properties
of the TBG model. The phase diagram includes the quantum anomalous
Hall and charge density wave phases at low $T$, and a Ising transition
separating them from the high-$T$ symmetric phases. Due to the proliferation
of excitons -- particle-hole bound states -- the transitions take place at
a significantly reduced temperature than the mean-field estimation.
The exciton phase is accompanied with distinctive experimental
signatures in such as charge compressibilities and optical conductivities
close to the transition. Our work explains the smearing of the
many-electron state topology by proliferating excitons
and opens the avenue for controlled many-body investigations on
finite-temperature states in the TBG and other quantum moir\'e systems.
\end{abstract}

\date{\today }
\maketitle

\textit{Introduction.---} Since the discovery of correlated insulators and 
superconducting states in the magic-angle twisted bilayer graphene 
(TBG) ~\cite{cao2018correlated,cao2018unconventional}, the 
quantum moir\'e systems \cite{Trambly2010,trambly2012numerical,
bistritzer2011moire} become an active playground for experimental 
and theoretical investigations on exotic phenomena
\cite{Yankowitz2019tbg,lu2019tbg,sharpe2019emergent,
chen2020tunable,kerelsky2019maximized,tomarken2019electronic,
xie2019spectroscopic,Shen2020DTBG,Serlin2020Science,Nuckolls_2020,
pierce2021unconventional,moriyama2019,Rozen2021Entropic,
Saito2021Pomeranchuk,liu2020spectroscopy,Naik2018,Conte2019,
Tang2020,Xian2020,Pan2020,Venkateswarlu2020,Lu2020,Zhang2020,Maity2021,
XuZhang2021,GPPan2021,XuZhang2021SC,TMDQAH2021,HQLi2021,YMXie2021}. 
The intriguing quantum effects in TBG and transition metal dichacogenide 
(TMD), besides the superconductivity, encompass a wide range including the
orbital ferromagnetism~\cite{sharpe2019emergent,Serlin2020Science},
(emergent) quantum anomalous Hall (QAH) effect \cite%
{Serlin2020Science,Nuckolls_2020,Wu2021NM,Das2021NP,TMDQAH2021}, large
(iso)spin entropy and Pomeranchuk effect~\cite%
{Saito2021Pomeranchuk, Rozen2021Entropic}, etc. Such a
plethora of quantum states is believed to originate from the interplay of strong electron correlation and fragile band topology~\cite{Po2018,
liu2018pseudo,po2018origin,fragile_topology,liu2019correlated,
LiuValley2019,LiuAnomalous2020,bultinck2019ground,Bultinck2020PRL,
TBGI_Bernevig2021,TBGII_Song2021,TBGIII_Bernevig2021,
TBGIV_Lian2021,TBGV_Bernevig2021,TBGVI_Xie2021}. Many theoretical models
have been put forward to address these interesting phenomena, and among the
positive semidefinite Hamiltonians that are proposed \cite%
{TBGIII_Bernevig2021,bultinck2019ground,kang2019strong,XuZhang2021Sign}, a real-space
effective model by Kang and Vafek (KV) is able to integrate both key ingredients~%
\cite{kang2018symmetry,kang2019strong}. The KV model and its
variants have been investigated via quantum Monte Carlo
simulations, which showed certain correlated insulating states such as the
inter-valley coherent and quantum valley Hall states, etc, are natural
ground states at integer fillings and particularly the charge neutral point
(CNP) of the TBG systems \cite%
{XYXu2018,YDLiao2019,YDLiao2020,YDLiao2020review}.

\begin{figure}[!t]
\includegraphics[angle=0,width=1\linewidth]{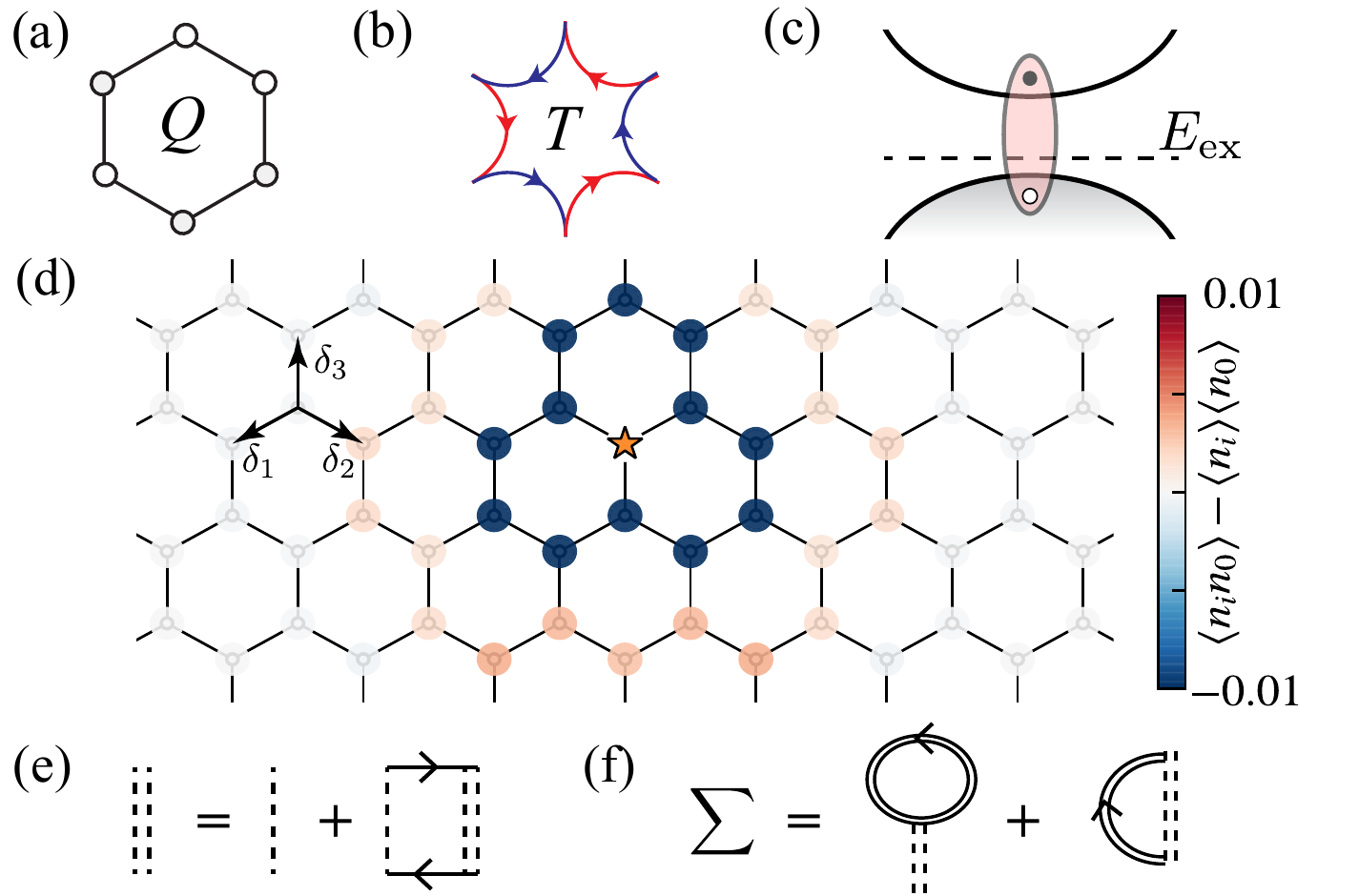}
\caption{The KV 
model with (a) the cluster charge $Q$ and
(b) assisted hopping $T$ terms. (c) illustrates the formation of
exciton between two quasi-particles from the
valence and conduction bands. (d) shows the charge correlations between a
central site (the asterisk) and other sites, at intermediate
temperature $T=0.244$ inside the exciton proliferation regime. 
{ The vectors $\delta_i$ are introduced to display the charge correlations in \Fig{Fig:ThermoQuant}(d).}
(e) shows the
scattering $T$-matrix that is the renormalized interaction of
electrons by the contributions of the ladder diagrams. (f) shows the
Hartree-Fock-like contribution to the self-energy of the single-particle
Green's function.}
\label{Fig:Model}
\end{figure}

Away from CNP, density matrix renormalization
group (DMRG) simulations upon the KV model revealed
that a QAH state can emerge purely from interactions at 3/4 filling~\cite{Chen2020TMI},
which constitutes the long-sought-after topological Mott insulator (TMI) \cite%
{topMott2008}. Along with other scenarios \cite%
{Bultinck2020PRL,Liu2021QAH,TBGIV_Lian2021} that give rise to the Chern
insulators at odd fillings, such a \textit{bona fide} TMI state
provides a strong coupling explanation of the
observed quantized Hall conductance in experiments \cite%
{Serlin2020Science,Nuckolls_2020}. Although the ground state with
non-trivial topology in the TBG model is known, the low-energy collective
excitations and the experimentally relevant finite-$T$ phase diagram are
still absent. As the QAH state spontaneously breaks time reversal
symmetry (TRS), a thermal phase
transition is expected to take place between the topological QAH and high-$T$ symmetric
phase. However, a naive estimate of the transition temperature $T_c$
according to the band gap \cite{Chen2020TMI} --- based on the mean field
theory --- leads to a value of the order of 100 K, higher than the
experimental value ($\lesssim 10$ K) by an order of magnitude~\cite%
{sharpe2019emergent,Serlin2020Science}. The difference is believed to stem
from the intertwinement of electronic interaction and thermal fluctuaions.

To elucidate the thermal melting of the topological phase and the
associated phase diagram, we perform accurate finite-$T$ many-body
calculations with the exponential tensor renormalization group (XTRG)
method~\cite{Chen2018,Lih2019}. XTRG calculations uncover an Ising-type
thermal phase transition between the low-$T$ QAH phase and the
symmetric phase, and a critical temperature reduced by one order
of magnitude compared to the mean-field estimation.
These observations are further explained by a field-theoretical approach,
which unveiled the emergence of a collective mode of a bounded particle-hole
pair--exciton. The excitons are found to have a rather flat and low lying dispersion
and proliferate as the temperature elevates, playing an essential role in
the melting of the low-$T$ phase.
The proliferation of excitons leads to a modulation of
electron-hole correlations in real space. 
The entire finite-$T$ phase diagram including the low-$T$ QAH and CDW
phases and the intermediate exciton proliferation phase, is accurately mapped
out via both XTRG and field-theoretical approaches.

These results extend the understanding of the zero-temperature phase
diagram to finite-temperature and dynamical effects of collective excitations
\cite{GPPan2021,KhalafSoftmodes2020}, beyond the exactly solvable limits
\cite{TBGIV_Lian2021,TBGV_Bernevig2021,KhalafSoftmodes2020,
kang2019strong,Vafek2021,Kwan2021}.  {Distinctive signatures
of the exciton proliferation phase are also revealed to bridge the experimentally
accessible observables with theoretical understanding.}

\begin{figure}[!t]
\includegraphics[angle=0,width=1\linewidth]{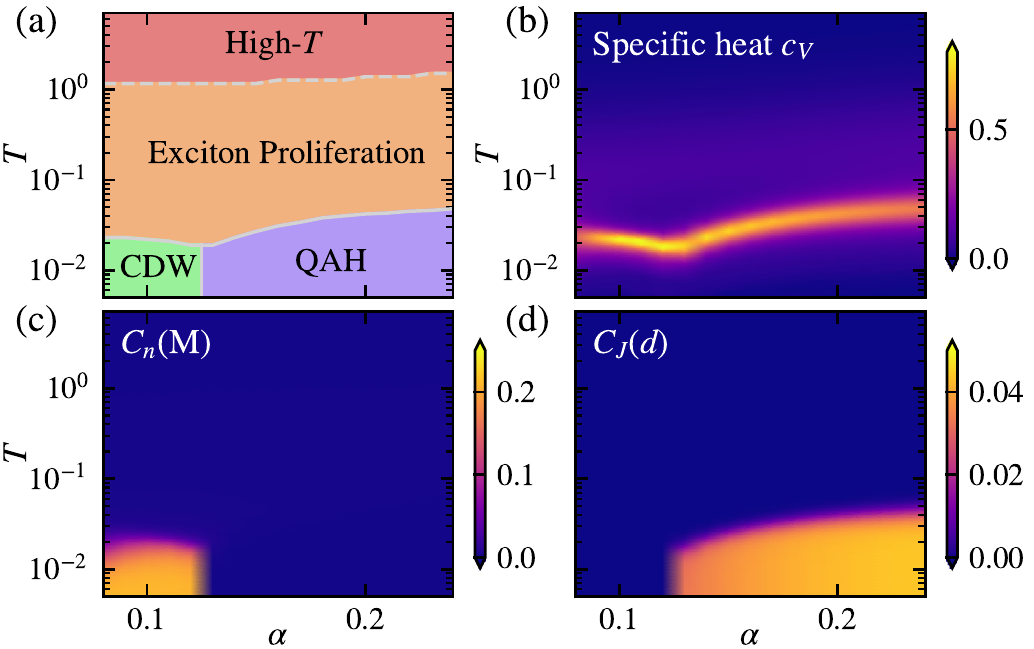}
\caption{ (a) Finite-temperature phase diagram of the KV model, with the red
regime being the high-$T$ disorder phase, the intermediate-$T$ orange one
where exciton proliferates, the green one being the CDW phase, and the
purple one being the QAH phase. The phase transition line between the low-$T$
symmetry breaking and the symmetric phase at intermediate $T$, as well as
the crossover temperatures (white dashed line) between intermediate- to the
high-$T$ regimes are determined from analyzing the specific heat $c_V$ data. 
Supporting the phase diagram in panel (a), we map the $%
\protect\alpha$-$T$ landscapes of (b) specific heat $c_V$, (c) charge
structure factor $C_n(\mathrm{M})$, and (d) current correlation $%
C_J(d)$ at a distance of $d=6$ to clearly identify the phase boundaries. }
\label{Fig:PhaseDiag}
\end{figure}

\begin{figure*}[!t]
\includegraphics[angle=0,width=1\linewidth]{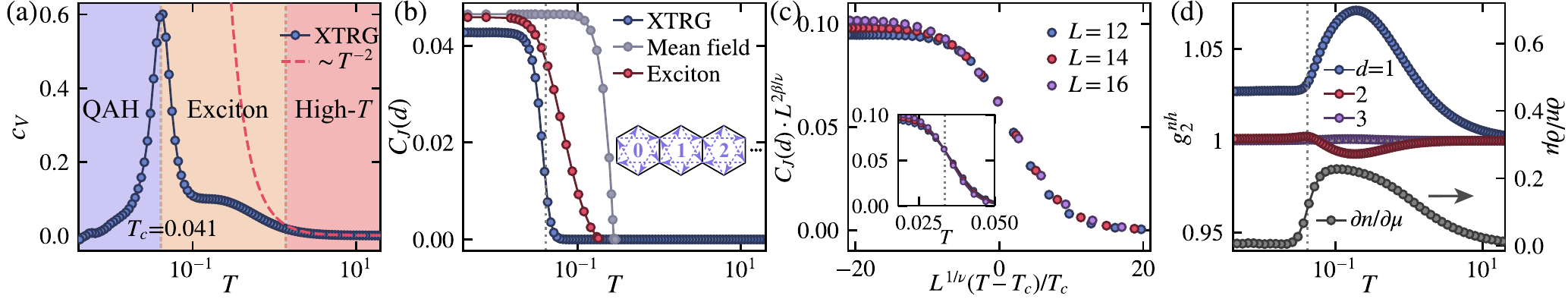}
\caption{ For the case of $\alpha=0.2$,
(a) shows the specific heat $c_V$ data, where the peak at $T_c=0.041$ signifies the
thermal phase transition. (b) Current-current correlation at a distance of $%
d=6$, i.e. $C_J(d=6)$ vs. $T$. Here, the
XTRG and mean-field, as well as the field-theoretical results (Exciton) are
shown. (c) Data collapse of the scaled current correlation $C_J(d=6)$ versus
$L^{1/\protect\nu}(T-T_c)/T_c$ with the Ising critical exponents $\protect%
\beta=\frac{1}{8}$ and $\protect\nu=1$, for the various system sizes. Insets
show the $C \cdot L^{2\protect\beta/\protect\nu}$ versus temperature $T$,
where the crossing point between different curves signifies
the transition temperature $T_c$.
{The left axis of panel }(d) shows the particle-hole correlation $g_2^{nh}(\mathbf{r})$
correlations with the reference site fixed at the center of the system
denoted as the asterisk and
{$\mathbf{r} = d(\mathbf{\delta_1}-\mathbf{\delta_2})$} denoted in 
{\Fig{Fig:Model}(d)}.
{The right axis of panel (d) shows the compressibility $\partial n/\partial\mu$ vs. $T$.}
The dashed lines in Panels (b,d) indicate the
specific heat peak $T_c=0.041$. }
\label{Fig:ThermoQuant}
\end{figure*}

\textit{Real-space TBG model, thermal tensor networks, and field-theoretical
approach.}--- The KV model considered here is described by
the interaction-only Hamiltonian
\begin{equation}  \label{Eq:H}
H = U_0\sum_{\varhexagon}({Q_{\varhexagon}} + \alpha T_{\varhexagon} - 1)^2,
\end{equation}
where 
$U_0 = 1$ sets the energy unit
{($\sim$ 40 meV in realistic system~\cite{Chen2020TMI}), the cluster
charge term $Q_{\varhexagon} \equiv \frac{1}{3}\sum_{l\in\varhexagon}
c^\dagger_l c^{\phantom{\dagger}}_l$ counts the numbers of electrons in each
hexagon [c.f., Fig.~\ref{Fig:Model}(a)], and $T_{\varhexagon} \equiv
\sum_{l\in\varhexagon} [(-1)^l c^\dagger_l c^{\phantom{\dagger}}_{l+1} +
\text{h.c.}]$ represents the assisted hopping term with alternating sign
[c.f. Fig.~\ref{Fig:Model}(b)]. The results of the present work are
mainly based on the YC4$\times L$ geometry with $L=12$, where the lattice is
under periodic/open boundary condition along vertical/horizontal direction
[c.f. Fig.~\ref{Fig:Model}(d) for a typical YC4 geoemtry]. We focus on the
3/4 filling of the TBG flat bands with projected Coulomb interaction and
correspondingly in Eq.~\eqref{Eq:H} this means the electron number $\langle
n_l (\equiv c^\dagger_l c^{\phantom{\dagger}}_l ) \rangle =1/2$ with the
valley and spin degrees of freedom polarized. Ref.~\cite{Chen2020TMI} identified a CDW insulator and a topologically
nontrivial QAH insulator in the ground state of the model, separated by a
first-order transition at $\alpha\simeq0.12$.}

Here, we explore the finite-$T$ properties of the TBG model with the XTRG method \cite%
{Chen2018,Lih2019},
which constitutes an accurate many-body method at finite
temperature, previously applied to simulate frustrated quantum magnets~\cite%
{Chen2018b,Lih2020,Li2021nc} as well as correlated fermions at both half
filling and finite doping~\cite{Chen2021}.
We retain up to $D=1000$ states, which renders the truncation errors $%
\delta<10^{-4}$ down to low-$T$ regime.

We also employ the Gaussian state theory and field-theoretical approach to
obtain both thermal and dynamical properties. The thermal state is approximated by a Gaussian ansatz, i.e., the optimal mean-field state, where the order parameters (i.e., all the single particle correlation functions) minimizing the free energy can be obtained efficiently via a set of flow equations \cite{shi2018,shi2020}. The particle-hole excitation spectrum can be obtained via the fluctuation analysis \cite{Guaita2019}, or alternatively the analytic structure of the scattering $T$-matrix, i.e., the renormalized electronic interaction [c.f. the ladder diagram in \Fig{Fig:Model}(e)]. As the temperature increases, the interaction is strongly renormalized, which drastically modifies the self-energy [c.f. the Hartree-Fock-like contribution in \Fig{Fig:Model}(f)] of the single particle Green function. More details on methodologies are presented
in the Supplemental Materials (SM)~\cite{suppl}.

\textit{Finite-temperature phase diagram.}--- Fig.~\ref{Fig:PhaseDiag}(a)
summarizes our finite-$T$ phase diagram, where the
low-$T$ phases include the CDW and QAH states, and the symmetric phase can
be divided into two regimes, with the orange intermediate-$T$ one acquring pronounced collective (exciton) excitations and the red high-$T$ state being trivial. As we will show below, there is a
crossover between the intermediate-$T$ and high-$T$ regimes, while the
intermediate-$T$ and low-$T$ phases are separated by a second-order phase
transition of Ising universality. Moreover, the two low-$T$ phases,
i.e., the CDW and QAH ones, are separated by a first-order transition extending from the transition point $\alpha\simeq0.12$ found in previous
DMRG study~\cite{Chen2020TMI}. %

In Fig.~\ref{Fig:ThermoQuant}(a), the specific heat $c_V$ shows
pronounced peaks around $T_c \simeq 0.041$ for $\alpha=0.2$, indicating the
existence of second-order phase transitions. The lower phase boundaries in
Fig.~\ref{Fig:PhaseDiag}(b)
are drawn in this way by performing different $\alpha$ scans.
The crossover line between the intermediate-
and high-$T$ regime is estimated by
collecting the temperature where the $c_V$ curve starts to scale as the
high-$T$ limit of $T^{-2}$ [c.f. the red dashed line in Fig.~\ref%
{Fig:ThermoQuant}(a)]~\cite{suppl}.

\textit{Melting of the QAH and CDW states.}---\ To detect
the QAH phase, we calculate the current correlation,
$C_J(d) = \langle J_0 J_d \rangle$,
where $J_0 \equiv \frac{1}{6}\sum_{l\in\varhexagon_0} \mathrm{i} (c^\dag_l
c^{\phantom{\dag}}_{l+2} - \mathrm{H.c.})/2$ is the average of all the
next-nearest-neighbor (NNN) currents [c.f. purple dashed arrows in the inset
of Fig.~\ref{Fig:ThermoQuant}(b)] inside the selected hexagon $\varhexagon_0$
in the $L/4$-th column of the cylinder, and $\varhexagon_d$ is the $d$-th
hexagon to the right of $\varhexagon_0$. We find $%
C_J(d) $ is nearly a constant in $d$ inside the QAH state \cite{suppl},
indicating the existence of a long-range order. For $T>T_c$, $C_J(d)$ decays exponentially with $d$. In Fig.~\ref{Fig:ThermoQuant}(b) we show the current correlation at a
sufficiently long distance 
in the QAH state, i.e., $C_J(d=6)$ versus $%
T $, which quickly drops and indicates the vanishing of QAH order above
the $c_V$ peak $T_c\simeq0.041$.
Notably, the mean-field results [depicted as the grey
dots] overestimate the transition temperature $T_c$ while the field
theoretical calculation including the contribution of exciton to the
quasiparticle self-energy, gives rise to $C_J(d)$ closer to the XTRG results,
and together they agree with the
corresponding temperature scale observed in the TBG experiment on the
melting of the QAH order~\cite{sharpe2019emergent,Serlin2020Science}. By
repeating the calculations of $C_J(d)$ for a wide range of $\alpha$, we
identify the existence of QAH phase with $\alpha>\alpha_c\simeq0.12$ in the $%
T$-$\alpha$ plane as shown in Fig.~\ref{Fig:PhaseDiag}(d).
{
Following the similar procedure, but using the charge structure factor
$C_n(\mathrm{M})\equiv\frac{1}{N}\sum_i e^{-i \mathrm{M}\cdot r_{0i}} (\langle \hat n_i
\hat n_0 \rangle - \langle\hat n_i\rangle\langle\hat n_0\rangle)$ at
$\mathrm{M}\equiv(0, \pm \frac{2\pi} {\sqrt{3}})$, we identify the CDW
phase in the left-bottom corner of the $T$-$\alpha$ plane, as shown in \Fig{Fig:PhaseDiag}(c).
}

Next, we address the universality class of the phase transition by finite-size data collapsing of the
current correlation $C_J(t,L)$ at $d=6$
with $t\equiv(T-T_c)/T_c$. In the vicinity of the critical point, we have $%
C(t,L)=L^{-2\beta/\nu}g(tL^{1/\nu})$, where the scaling function behaves $%
g(x)\sim x^{\beta}$ as $x\rightarrow\infty$.
As the low-$T$ phase breaks $Z_2$ (TRS) 
symmetry, it is natural to expect a
2D Ising universality class. To verify this, the critical exponents $%
\beta=1/8$ and $\nu=1$  of the order parameter and correlation
length, respectively, are used to collapse the finite-$T$ data in Fig.~\ref%
{Fig:ThermoQuant}(c). In the inset of Fig.~\ref{Fig:ThermoQuant}(c), we
plotted the $C_J L^{2\beta/L}$ versus $T$ and identify the critical
temperature $T_c\simeq0.034$ as the crossing points between curves of
different system sizes. This is the $T_c$ at the thermodynamic limit and slightly different from
the peak in the specific peak in Fig.~\ref{Fig:ThermoQuant}(a) for one system size. 
Then we rescale the $x$-axes as $t L^{1/\nu}$ and thus see perfect data collapses
within the critical regimes in the main panels of Fig.~\ref{Fig:ThermoQuant} (c).

\textit{Electron-hole correlation and exciton proliferation.}--- As
temperature rises above $T_c$, the excitons proliferate and result in
nontrivial features on the charge correlations. As shown in Fig.~\ref%
{Fig:Model}(d) with $T=0.244$ and $\alpha=0.2$, we place a hole at the very
center of the lattice, and find electrons exhibit bunching and anti-bunching
modulation behaviors as moving away from the hole, evidencing the existence
of particle-hole bound states --- excitons --- in the system. Such peculiar
charge correlations can be quantitatively reflected in the electron-hole
correlator $g_2^{nh}({d}) = \frac{\langle \hat n_0 \hat h_i \rangle}{%
\langle \hat n_0\rangle \langle \hat h_i \rangle}$, with $\hat h_i=1-\hat
n_i $ and measured between two sites (``0" and ``$i$") separated by
distances {$\mathbf{r}=d(\mathbf{\delta_1-\delta_2})$} [c.f. Fig.~\ref{Fig:ThermoQuant}(d)], 
which show increasing correlation whose ``sign'' changes for different distance in the 
regime $T>T_c$, and shows extremum values at an intermediate temperature $T_\mathrm{ex}\simeq0.3$ 
(around the mean-field transition temperature), at which the excitons can be easily excited 
since the single-particle gap 
now roughly equals the thermal energy scale. The electrons 
and holes at nearest sites belonging to the same sublattice
[i.e., connected via {$(\mathbf{\delta_1}-\mathbf{\delta_2})$} as shown in 
{Fig.~\ref{Fig:Model} (d)}] attract each other, while at further distance like 
{$d=2$} they repeal, reflecting the strong influence of
excitons in the thermal states. As distance further enhances, e.g., 
{$d=3$,} the electron-hole correlations become rather
weak, showing that the excitons are indeed quite local. 
{We note that, the charge correlation decays exponentially with distance, 
and thus the oscillation behavior is not attributed to Friedel oscillations.}
When the temperature further elevates and goes beyond {$T_\mathrm{ex}$}, 
even the short-range charge correlations get smeared by strong thermal
fluctuations, all correlations decay ($g_2^{nh} \to 1$) at about {$T_h \simeq 1$}.
Above this crossover temperature, i.e., in the high-$T$ regime, specific heat 
$c_V$ exhibit $\sim 1/T^2$ scaling as illustrated in Fig.~\ref{Fig:ThermoQuant}(a) ~\cite{suppl}.

{
We also {study the compressibility} $\partial n/\partial\mu$
by adding a chemical potential term to { the} KV model \cite{suppl}. 
In \Fig{Fig:ThermoQuant}(d), the compressibility exhibits
a steep jump above $T_c$ and keeps an enhanced value inside the exciton
regime. This is a direct {result of the exciton proliferation above $T_c$, where the formation of excitons (bosonic bound state) significantly
enhanced the compressibility.} Such a steep enhancement can be measured in the quantum capacitance and
scanning single electron transistor experiments~\cite{Eisenstein1992,
wong2019cascade,zondiner2019cascade}. In fact, the compressibility enhancement above
the correlated insulators (CDW and QAH phases), is qualitatively consistent with the experimental observation
at the same $3/4$ filling of TBG~\cite{zondiner2019cascade}. }

\begin{figure}[!t]
\includegraphics[angle=0,width=1\linewidth]{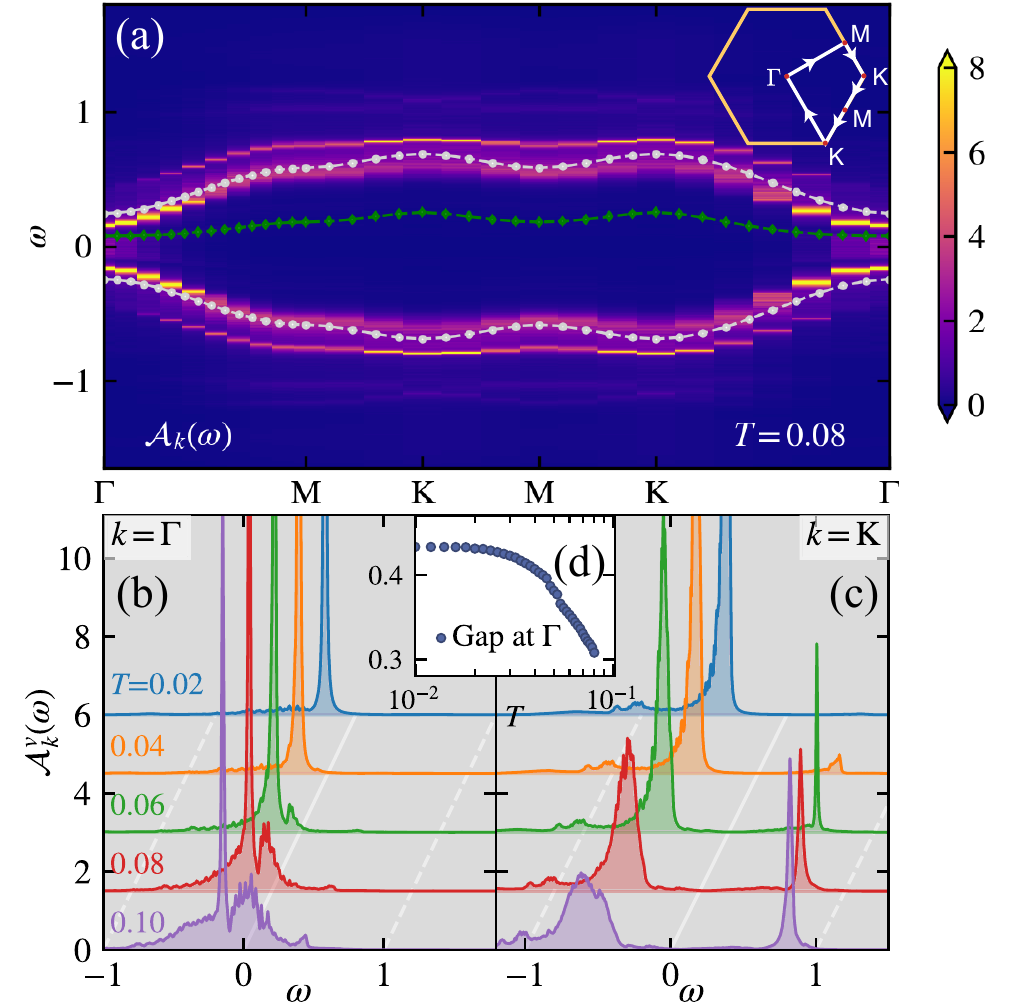}
\caption{(a) Quasi-particle spectral function
$\mathcal{A}_k(\omega)=\mathcal{A}_k^v(\omega)
+\mathcal{A}_k^c(\omega)$ as the
sum of valence and conducting band spectrum,
for $\alpha=0.2$ at $T=0.08$,
where the spectral weight distribution gets broadened as compared to
the mean-field single-particle dispersion consisted of upper conducting
and lower valence bands (white dots). The exciton band lying within
the single-particle gap is denoted by green diamonds.
The inset depicts the path in the Brillouin zone.
Panels (b,c) show the valence electron spectral weights at
$k=\Gamma$ and $k=\mathrm{K}$ for various temperatures,
where the values of $\mathcal{A}$ and $\omega$ are properly
shifted as indicated by the tilted white lines.
(d) shows the gap, defined as twice the peak position of $\mathcal{A}_\Gamma^v(\omega)$,
as a function of temperature $T$.
}
\label{Fig:CorFunc}
\end{figure}

\textit{Dynamical signature of excitons.}--- At the mean-field level, the
gap between the conductive and valence bands (white dots in Fig.~\ref{Fig:CorFunc} (a)) in the QAH phase is about $0.5 U_0$ (at the $\Gamma$ point in BZ), giving rise to a transition to the disorder phase at
much higher temperature $T_c \sim 0.2$ (of the scale of 100 K for realistic materials), at the scale of the ground state band gap~\cite{Chen2020TMI}.
However, our XTRG computation finds a much lower transition temperature $T_c \sim 0.04$ ($\sim$ 10 K) by one order of magnitude, which agrees with the experimental results
\cite{sharpe2019emergent,Serlin2020Science}
implying the failure of the mean-field theory at finite temperatures.

To explore the mechanism of such low transition temperature, we perform the diagrammatic calculation (see details
in SM~\cite{suppl}) on the $T$-matrix in the particle-hole channel and its correction to the self-energy of single-particle Green functions. The poles of the $T$-matrix determine the exciton spectrum. As shown by the green diamonds in Fig.~\ref{Fig:CorFunc} (a) at a representative temperature $T=0.08$, the exciton has a much lower energy $E_\mathrm{ex} \sim 0.08$ than the mean-field gap. As a result, at the finite temperature comparable with $E_\mathrm{ex}$, many excitons are proliferated by thermal fluctuations, and the scattering with excitons strongly affects single electron behaviors in conductive and valence bands. The effect of excitons to the single particle Green function can be characterized by the $T$-matrix in the Hartree-Fock correction to the self energy (Fig.~\ref{Fig:Model}(e) and (f)). In Fig.~\ref{Fig:CorFunc} (a), we show such renormalized spectral function $A_\mathbf{k}^{c,v}(\omega)$ for electrons in conductive and valence bands at $T=0.08$, which displays much smaller band gap than that from the mean-field theory.
In Fig.~\ref{Fig:CorFunc}(b,c), the spectral functions of electrons in the valence band at different temperatures show the reduced quasi-particle weight and the broadened peak as $T$ increases.
It is remarkable that at the temperature inside the exciton regime, the collective mode {assists the valence electron to tunnel} across the band gap and {redistribute} in the positive frequency region. This intriguing feature is the signature of the electron dressed by the cloud of proliferated excitons,
{
which can be probed via spectroscopy. The absorption spectra of probe light display peaks
at the exciton frequencies ($\sim 100$ GHz shown in \Fig{Fig:CorFunc}).}
Due to the strong exciton dressing, the current correlation function is thus highly reduced, as shown by the red dots in Fig.~\ref{Fig:ThermoQuant} (b). Accompanied with the results in Figs.~\ref{Fig:Model} (d) and ~\ref{Fig:ThermoQuant} (d), our results provide {a direct observation of proliferated excitations} in the 3/4 filling
setting of TBG model.
{
We also confirm, a small kinetic term will not qualitatively
change the exciton physics observed here \cite{suppl}.}

\textit{Discussion.}--- We extend the studies of the TBG model to
the finite-temperature properties and collective excitations. In particular,
the excitons formed by a pair of quasi-particle and hole proliferate at 
intermediate temperatures, which significantly influences the charge 
correlations and provide a mechanism to melt the QAH phase. We 
therefore reconcile the large quasi-particle band gap and small QAH 
transition temperature observed in experiments. The excitons, 
as collective excitations due to Coulomb interactions, have been 
investigated and discussed in bilayer graphene~\cite{Ju2017,Wu2017}, 
TBG~\cite{Kwan2021,Liu2017}, as well as the recent observation of the 
QAH in TMD heterobilayers~\cite{TMDQAH2021}. { We note 
that, the exciton physics here is a distinct feature of flat-band system, 
different from the Haldane-Hubbard model (see, e.g., Ref.~\cite{CanShao2021}) 
where the excitons comprised of band electrons and holes do not 
experience a proliferation upon rising temperature~\cite{suppl}.} 
Our work here shows the emergence 
of excitons in the strong coupling limit in a TBG lattice model, proposing 
intriguing exciton physics {such as the charge compressibility and the 
spectral fingerprint in both single-particle and collective models,} 
waiting to be explored in future experiments of quantum moir\'e systems. 
We point out that, the valley and spin degrees of freedom, which may 
give rise to other neutral collective modes (e.g. magnon) and possible 
Pomeranchuk physics, are omitted in the present model and will be 
addressed in a future study.

\begin{acknowledgments}
\textit{Acknowledgments.}--- X.Y.L. and B.B.C. contributed equally to this
work. X.Y.L., W.L., Z.Y.M. and T.S. are indebted to Jian Kang  and Yifan Qu for stimulating
discussions. X.Y.L., W.L. and T.S. acknowledge the support from the NSFC through Grant Nos.
11974036, 11834014, 11974363 and 12047503. B.B.C. and Z.Y.M. acknowledge support from
the RGC of Hong Kong SAR of China (Grant Nos.  17303019,
17301420, 17301721 and AoE/P-701/20), the Strategic
Priority Research Program of the Chinese Academy of
Sciences (Grant No. XDB33000000), the K. C. Wong
Education Foundation (Grant No. GJTD-2020-01) and
the Seed Funding "Quantum-Inspired explainable-AI" at
the HKU-TCL Joint Research Centre for Artificial Intelligence. We thank the High-performance Computing Center at ITP-CAS,
the Computational Initiative at the Faculty of Science and Information
Technology Service at the University of Hong Kong, and the Tianhe platforms
at the National Supercomputer Centers for their technical support and
generous allocation of CPU time.
\end{acknowledgments}

%


\newpage\clearpage
\renewcommand{\theequation}{S\arabic{equation}} \renewcommand{\thefigure}{S%
\arabic{figure}} \setcounter{equation}{0} \setcounter{figure}{0}

\begin{widetext}

\section{Supplemental Materials for \\[0.5em]
Exciton Proliferation and Fate of the Topological Mott Insulator in a
\newline
Twisted Bilayer Graphene Lattice Model}

\subsection{Section I: Exponential Tensor Renormalization Group Method}

\begin{figure}[!h]
\includegraphics[angle=0,width=.5\linewidth]{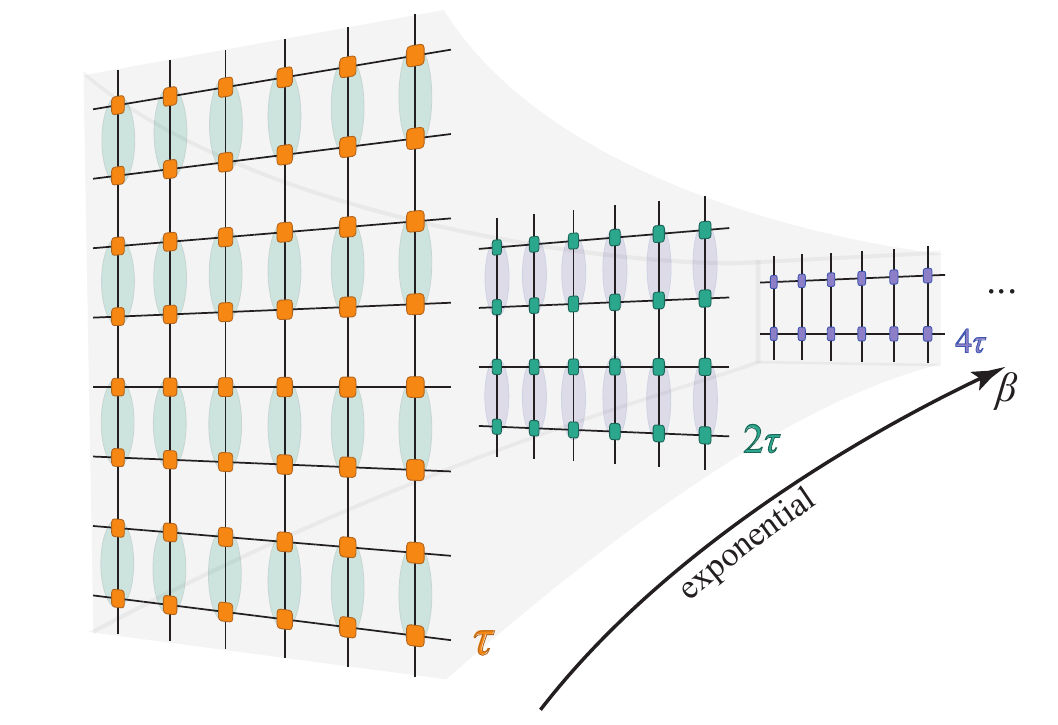}
\caption{Exponential evolution scheme in exponential tensor renormalization
group (XTRG) method.}
\label{FigS:XTRG}
\end{figure}

As shown in Fig.~\ref{FigS:XTRG}, the main idea of exponential tensor
renormalization group (XTRG) \cite{Chen2018,Lih2019} method is, to first
construct the initial high-temperature density operator $\hat\rho_0\equiv%
\hat\rho(\tau) = e^{-\tau \hat H}$ with $\tau$ being an exponentially small
inverse temperature, which can be obtained with ease via Trotter-Suzuki
decomposition \cite{Trotter1959} or series-expansion methods \cite{Chen2017}%
. Subsequently, we evolve the thermal state exponentially by squaring the
density operator iteratively, i.e., $\hat\rho_n\cdot\hat\rho_n\equiv\hat%
\rho(2^n\tau)\cdot\hat\rho(2^n\tau)\rightarrow\hat\rho_{n+1}$. Following
this exponential evolution scheme, one can significantly reduce the
evolution as well as truncation steps, and thus can obtain highly accurate
low-$T$ data in greatly improved efficiency. \vskip4mm In XTRG simulations,
we compute the internal energy per site
\begin{equation}  \label{Eq:En}
u(T)\equiv\frac{1}{N}\frac{\mathrm{Tr}[H\cdot\rho(T)]}{\mathrm{Tr}[\rho(T)]},
\end{equation}
where $H$ is the Hamiltonian [c.f. Eq.~({\color{blue}1})] and $\rho(T)\equiv
e^{-H/T}$ is the density matrix of the system with $N$ sites, and the
specific heat via the derivative of internal energy, is
\begin{equation}  \label{Eq:cV}
c_V(T) \equiv -\frac{\partial u(T)}{\partial T} = \frac{\partial u(\beta)}{%
\partial \ln\beta}\beta,
\end{equation}
with $\beta\equiv1/T$ the inverse temperature.

\vskip4mm

\begin{figure}[!h]
\includegraphics[width=.6\linewidth]{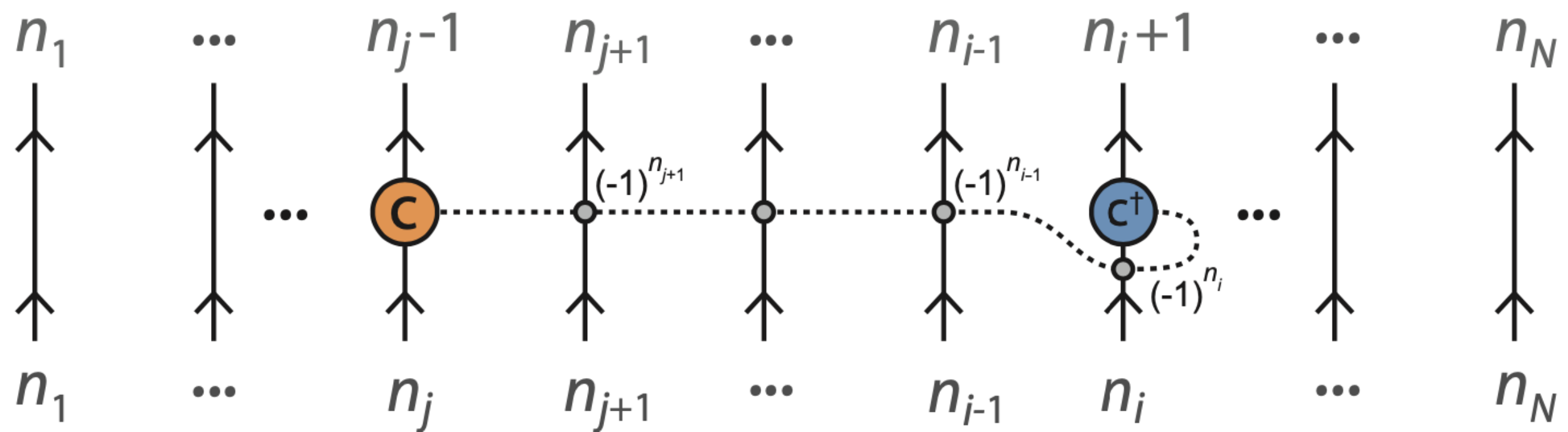}
\caption{Representation of one-body operator $c^\dagger_i c^{\,}_j$ as a
matrix product operator with bond dimenstion $D=1$.}
\label{FigS:XTRG-FermiSign}
\end{figure}

When adapting XTRG to fermion systems, one should take care of the fermionic
sign of exchanging two electrons. In this work, we are working on the
many-body basis $|n_1 n_2 \cdots
n_N\rangle\equiv(c_N^\dag)^{n_N}\cdots(c_2^\dag)^{n_2}(c_1^\dag)^{n_1}|%
\Omega\rangle$, where $n_i(\in\{0,1\})$ is the number of electrons at the
site $i$ and $|\Omega\rangle$ is the vacuum state. Generically in this
basis, the one-body operator $c_i^\dag c_j^{\phantom{\dag}}$ (assuming $j<i$%
) requires an sign $\Pi_{l=j+1}^{i}(-1)^{n_l}$, in addition to transform the
state $|n_1\cdots n_j \cdots n_i \cdots n_N\rangle$ to the state $|n_1\cdots
n_j\mathrm{-1} \cdots n_i\mathrm{+1} \cdots n_N\rangle$. As shown in Fig.~%
\ref{FigS:XTRG-FermiSign}, such fermion-sign structure can be encoded in
XTRG readily in a matrix product operator with bond dimension $D=1$.

\subsection{Section II: Crossover between exciton proliferation regime and
high-T regime}

\begin{figure}[!h]
\includegraphics[angle=0,width=.65\linewidth]{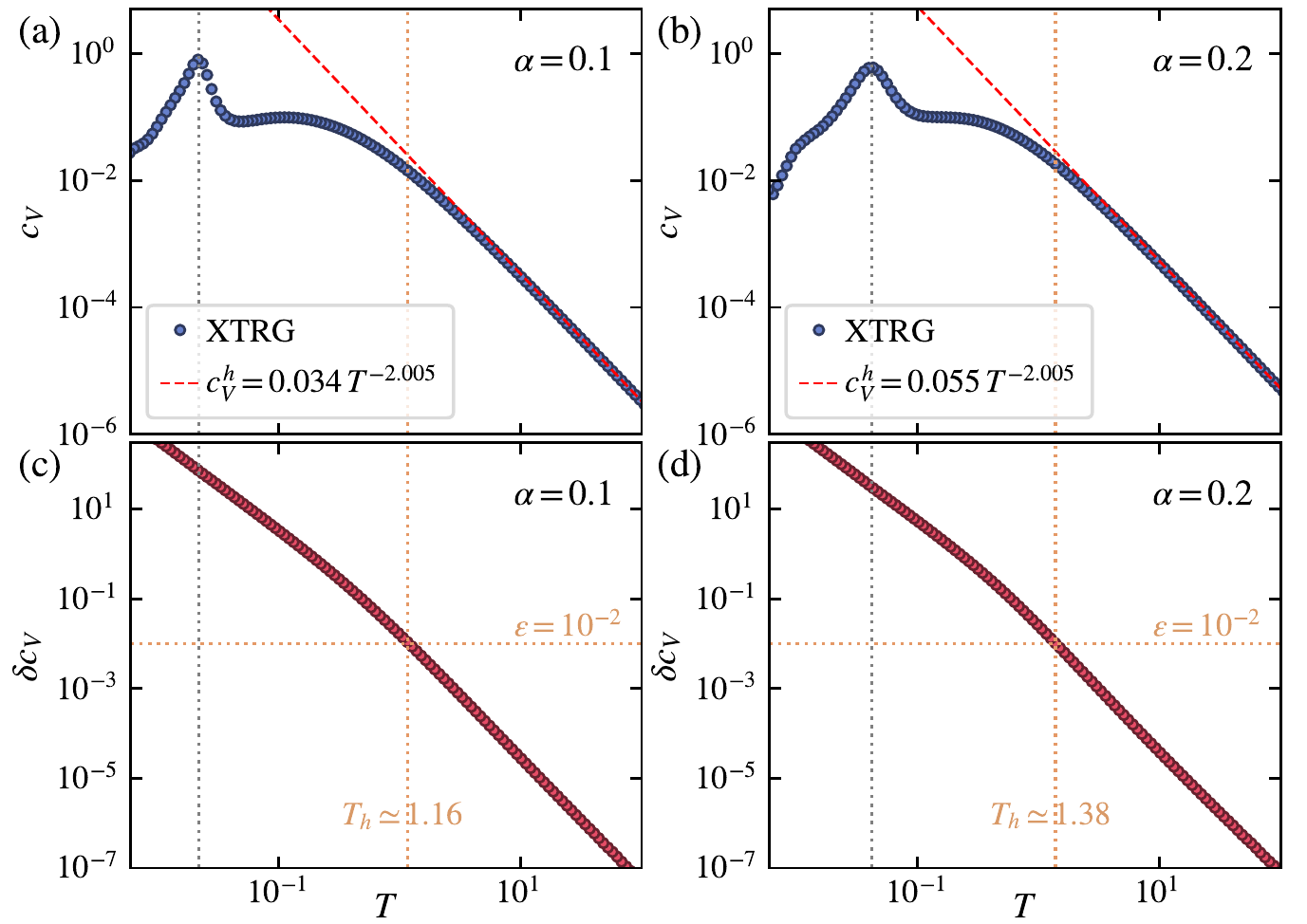}
\caption{In a YC4$\times$12 system for (a) $\protect\alpha=0.1$, and (b) $%
\protect\alpha=0.2$, specific heat $c_V$ curves are shown versus temperature
$T$ in a double-logarithmic scale. At high-temperature limit, $c_V$ behaves
as $\sim T^{-2}$, and is fitted by the red dashed line $c_V^h$. Panels (c)
and (d) show the deviation $\protect\delta c_V = |c_V - c_V^h|$ as functions
of temperature $T$. The crossover temperature between intermediate-$T$ regime
and the high-$T$ one is determined by a threshold $\protect\epsilon=10^{-2}$.}
\label{FigS:HighT}
\end{figure}

In this section, we will discuss the determination of the crossover between
the intermediate-$T$ exciton proliferation regime and the high-$T$ gas-like
regime. In a high temperature $T=1/\tau$ (i.e., a small inverse temperature $%
\tau$), the internal energy density of the system
\begin{equation*}
u(\tau) = \frac{1}{N}\frac{\mathrm{Tr}[H\cdot\rho(\tau)]}{\mathrm{Tr}%
[\rho(\tau)]} = \frac{1}{N}\frac{\mathrm{Tr}(H\cdot e^{-\tau H})}{\mathrm{Tr}%
(e^{-\tau H})},
\end{equation*}
via Taylor expansion, can be expressed as
\begin{equation*}
u(\tau) = \frac{1}{N}\left[ \frac{\mathrm{Tr}(H)}{Z^0} - \frac{\mathrm{Tr}%
(H^2)}{Z^0}\tau + \left(\frac{\mathrm{Tr}(H)}{Z^0}\right)^2\tau + O(\tau^2) %
\right].
\end{equation*}
Thus in the large-$T$ limit, it yields
\begin{equation*}
c_V\equiv\partial u/\partial T = (\partial u/\partial \tau) \cdot (\partial
\tau/\partial T) \sim T^{-2}
\end{equation*}
for specific heat.

As shown in Fig.~\ref{FigS:HighT}, we show the specific heat $c_V$ of a YC4$%
\times$12 system at both $\alpha=0.1$ and $\alpha=0.2$, which show
predominant power-law decay at high-T regime. As indicated by the red dashed
lines in Fig.~\ref{FigS:HighT}(a,b), we find the high-$T$ data
asymptotically follow $c^h_V = 0.034 T^{-2.005}$ and $c^h_V = 0.055
T^{-2.005}$ for $\alpha=0.1$ and $\alpha=0.2$ respectively, which are well
consistent with the large-$T$ limit. We thus determine the crossover
temperature $T_h$ between the high-$T$ regime and intermediate-$T$ regime,
by computing the deviation of the specific heat from the high-T behavior,
i.e., $\delta c_V = |c_V - c_V^h|$. To be more specific, we classify those
temperatures at which $\delta c_V>\epsilon=10^{-2}$ as intermediate-$T$, and
otherwise as high-$T$.

\subsection{Section III: Detailed current-current correlation calculation}

\begin{figure}[!h]
\includegraphics[angle=0,width=.75\linewidth]{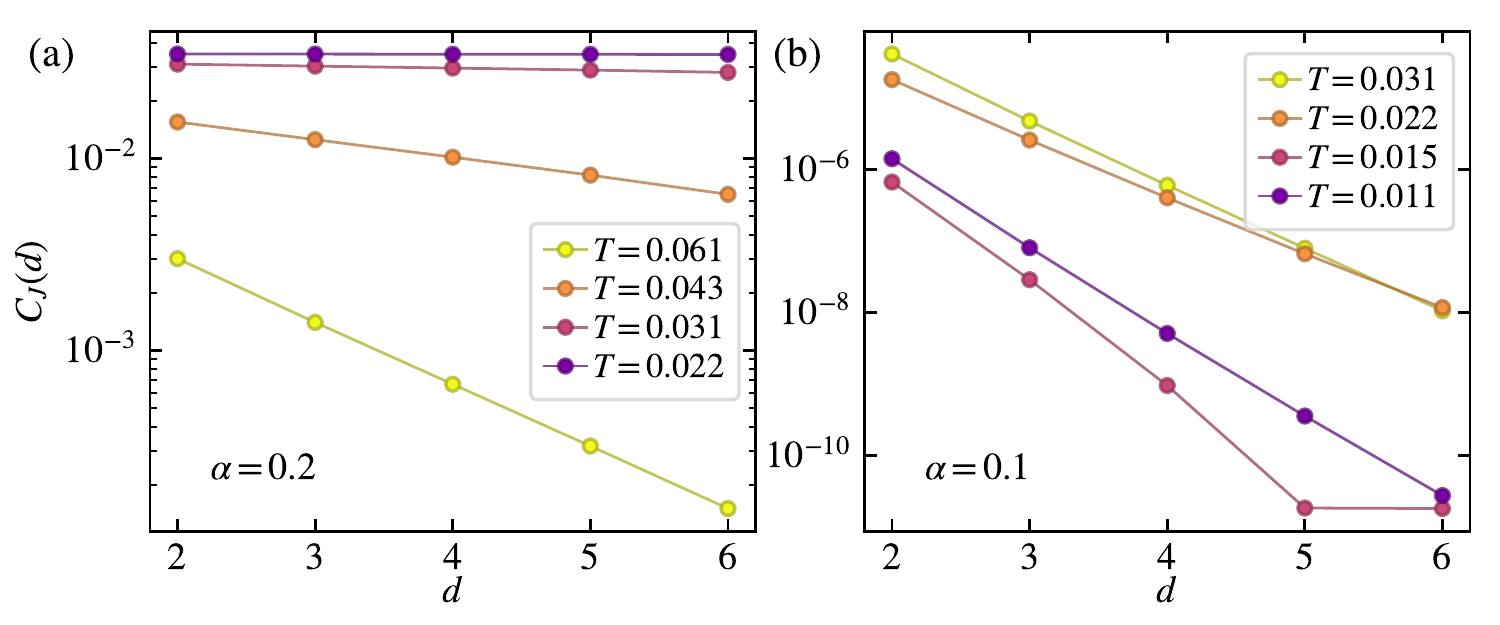}
\caption{In a YC4$\times$12 system for (a) $\protect\alpha=0.2$, and (b) $%
\protect\alpha=0.1$, the current-current correlation $C_J(d)$
is obtained versus distances $d$ between hexagons, for
various temperatures as indicated by the colored markers.}
\label{FigS:CjDetails}
\end{figure}

As shown in Fig.~\ref{FigS:CjDetails}, we calculate the current-current
correlation function $C_J(d)$, defined in the main
text. In a YC4$\times$12 systems at $\alpha=0.2$, $C_J$ establish a plateau
over distance $d$, in the quantum anomalous Hall (QAH) region, i.e., for $%
T<T_c\simeq0.041$, whereas it decays exponentially for $T>T_c$. It means
that, the lower-$T$ region for the large $\alpha$ cases spontaneously break
the time-reversal symmetry, manifesting the QAH state. On the other hand,
for the small-$\alpha$ case ($\alpha=0.1$ here), $C_J(d)$ decays
exponentially for both regions of $T<T_c\simeq0.022$ and $T>T_c$.

\subsection{Section IV: More details on the CDW data}

\begin{figure}[!h]
\includegraphics[angle=0,width=\linewidth]{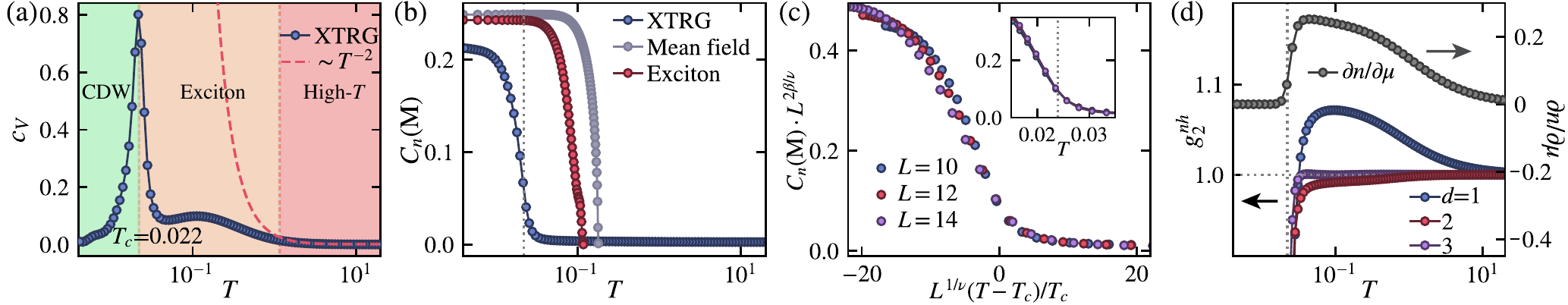}
\caption{ (a) The specific heat $c_V$ data are shown for $\protect\alpha=0.1$%
, where the peaks at $T_c=0.022$ signify the thermal phase transitions. (b)
Charge density structure factor $C_n(k) = \frac{1}{N}\sum_i e^{-ik\cdot
r_{0i}} (\langle \hat n_i \hat n_0 \rangle - \langle\hat
n_i\rangle\langle\hat n_0\rangle)$ with $k=\mathrm{M}$, vs. $T$ for the $%
\protect\alpha=0.1$ case. Here, the XTRG and mean-field results are shown.
(c) Data collapse of the scaled $C_n(\mathrm{M})$ versus $L^{1/\protect\nu%
}(T-T_c)/T_c$ with the Ising critical exponents $\protect\beta=\frac{1}{8}$
and $\protect\nu=1$, for the various system sizes. Insets show the $C \cdot
L^{2\protect\beta/\protect\nu}$ versus temperature $T$, where the crossing
point between curves of different system sizes signifies the thermodynamic limit transition
temperature $T_c=0.024$, slightly deviated from the specific heat peak at
finite length $L=12$.
{The left axis of panel}
(d) shows the particle-hole correlation $g_2^{nh}(%
\mathbf{r})$ correlations with the reference site fixed at the center of the
system [c.f. Fig.~{\color{blue}1}(d) of the main text].
{The right axis of panel (d) shows the compressibility $\partial n/\partial \mu$ vs. $T$.}
The dashed lines in Panels (b,d) indicate the specific heat peak $T_c=0.022$. }
\label{FigS:CDWData}
\end{figure}

In this section, we discuss more detailed thermodynamics results of the CDW
phase for $\alpha<\alpha_c$. In Fig.~\ref{FigS:CDWData}(a), we show the
specific heat $c_V$ curve vs. temperature $T$, which peaks at $%
T_c\simeq0.022 $ predominantly, indicating a phase transition there. We also
compute the density-density correlation function $\langle \hat n_i \hat n_0
\rangle\equiv\mathrm{Tr} (\hat\rho \,\hat n_i \hat n_0)$ at various
temperatures, with site $0$ being fixed at a center site and site $i$
running over the lattice. The charge structure factor
\begin{equation}
C_n(k)\equiv\frac{1}{N}\sum_i e^{-ik\cdot r_{0i}} (\langle \hat n_i \hat n_0
\rangle - \langle\hat n_i\rangle\langle\hat n_0\rangle),  \label{Eq:Cn}
\end{equation}
is found to peak at $\mathrm{M}\equiv(0, \pm \frac{2\pi} {\sqrt{3}})$ point
in the Brillouin zone (BZ). Note there are three pairs of equivalent $M$ points in the BZ
while only one of them is preferred by the cylindrical geometry, c.f., Ref.~\cite%
{Chen2020TMI}. As shown in Fig.~\ref{FigS:CDWData}(b), for the low-$T$ CDW
state, the CDW order parameter $C_n(\mathrm{M})$, structure factor at the $%
\mathrm{M}$ point, experiences a sudden drop at the transition temperature $%
T_c \simeq0.022$ upon heating, the same temperature as the specific heat
peak locates in Fig.~\ref{FigS:CDWData}(a). Again, the mean-field results
overestimate $T_c$, which {{is believed to be}}
corrected by the higher-order perturbative calculation towards the XTRG
results.

To address the universality class of phase transitions between the CDW phase
that breaks $Z_2$-type (discrete translational) symmetries to the symmetric
phase at higher temperatures, we follow the same line as in the main text
for the current-current correlation $C_J$, and perform the finite-size data
collapsing of $C_n(\mathrm{M})$ in Fig.~\ref{FigS:CDWData}(c). As a function
of $t$ ($\equiv\frac{T-T_c}{T_c}$) and system size $L$, we denote it as
$C_n(\mathrm{M};t,L)$.
We again using the 2D Ising critical exponents $\beta=1/8$ and $\nu=1$ as
the CDW phase breaks $Z_2$ symmetry. In the inset of Fig.~\ref{FigS:CDWData}%
(c), we plotted the $C_n(\mathrm{M};t,L) L^{2\beta/L}$ versus $T$ and
identify the critical temperature $T_c\simeq0.024$ as the crossing points
between curves of different system sizes. The so-estimated $T_c$ is {%
{again very closed to}} the peak of specific heat. Then
we rescale the $x$-axes of Fig.~\ref{FigS:CDWData}(c) as $t L^{1/\nu}$ and
see perfect data collapses within the critical regimes in the main panel of
Fig.~\ref{FigS:CDWData}(c).

In Fig.~\ref{FigS:CDWData}(d), we see that the $g_2^{{nh}}$ correlation
increases rapidly as the long-range anti-bunching correlation melts at the
CDW transition temperature. Other than that, we also observe similar
particle-hole correlation modulation in the exciton-proliferated
intermediate-$T$ regime above the CDW phase, as in Fig.~{\color{blue}3}%
(d) of the main text.
In \Fig{FigS:CDWData}(d), we also {perform the calculation on the compressibility} $\partial n/\partial\mu$
by adding a chemical potential term to {the KV model}. Similarly, the compressibility exhibits
a steep jump above $T_c$ and keeps an enhanced value inside the exciton
regime.

\subsection{Section V: Detailed $\protect\alpha$-scan for phase diagram}

\begin{figure}[!h]
\includegraphics[angle=0,width=\linewidth]{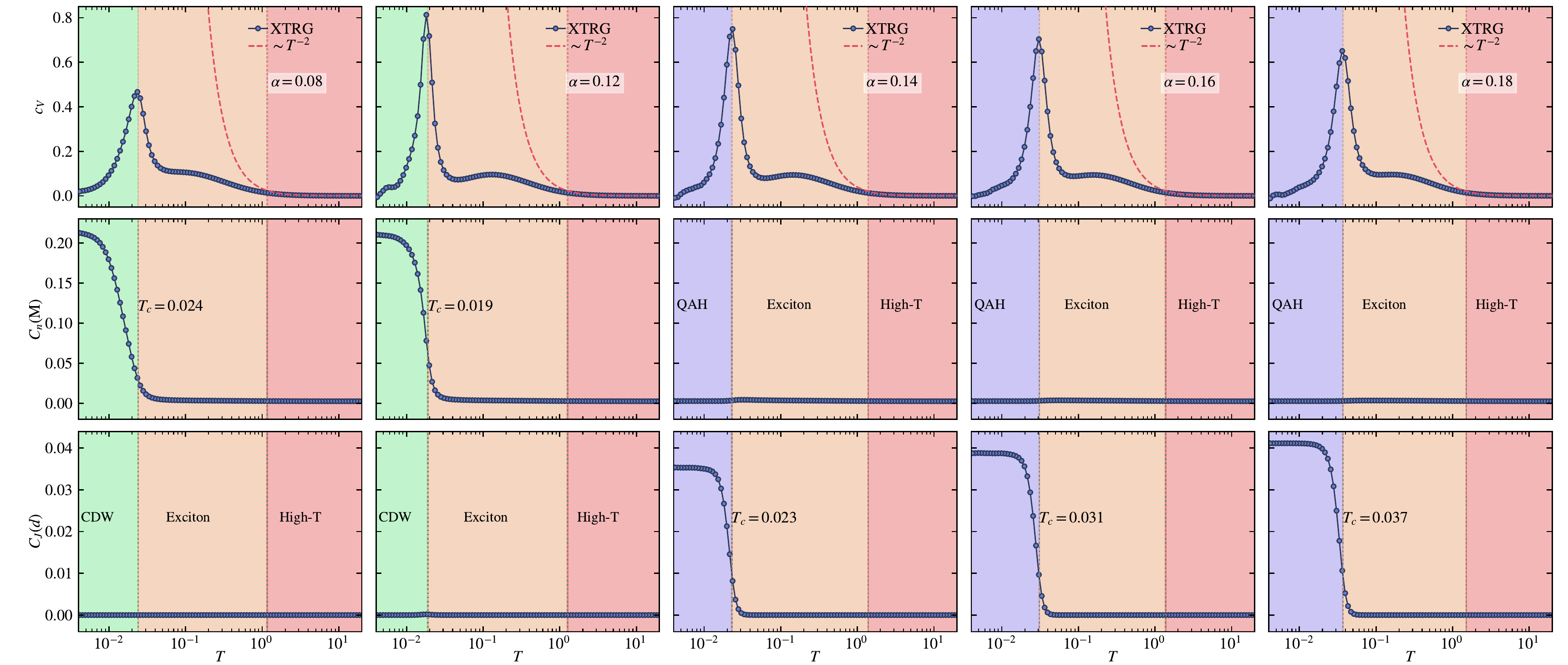}
\caption{In a YC4$\times$12 system for various $\protect\alpha=0.08, 0.12,
0.14, 0.16, 0.18$, the specific heat $c_V$ (the first row), the charge
density structure factor $C_n(\mathrm{M})$ (the second row), and the
current-current correlation $C_J(d=6)$ (the last row), are shown versus
temperature $T$.}
\label{FigS:AlphaScan}
\end{figure}

In this section, we will show more results of specific heat $c_V$, charge
structure factor $C_n(\mathrm{M})$, and the current-current correlation
function $C_J(d=6)$ for various $\alpha$ other than $\alpha=0.1$ and $0.2$,
as complement to the main text. As shown in Fig.~\ref{FigS:AlphaScan}, in
all these cases ($\alpha=0.08, 0.12, 0.14, 0.16, 0.18$) the specific heat $%
c_V$ curves (the first row) clearly show sharp peaks, above which either $%
C_n(\mathrm{M})$ ($\alpha=0.08$ and $0.12$) and $C_J(d=6)$ ($\alpha=0.14,
0.16$ and $0.18$) quickly vanish.

\subsection{Section VI: Electronic compressibility calculations}

\begin{figure}[htb]
\includegraphics[width=.75\linewidth]{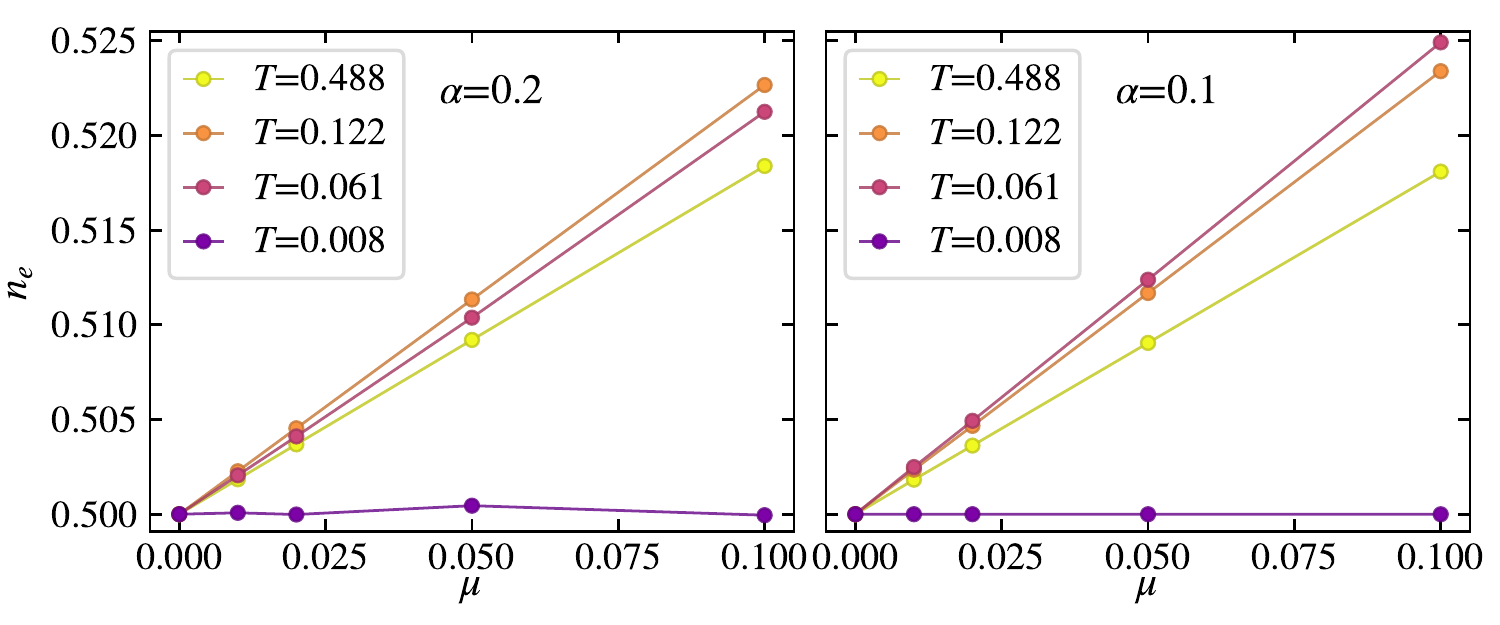}
\caption{The averaged charge density $n_e$ as a function of chemical potential $\mu$ at
various temperatures in the cases of $\alpha=0.2$ (Left panel) and $\alpha=0.1$ (Right panel).}
\label{FigS:ParticleNum}
\end{figure}

\begin{figure}[htb]
\includegraphics[width=.75\linewidth]{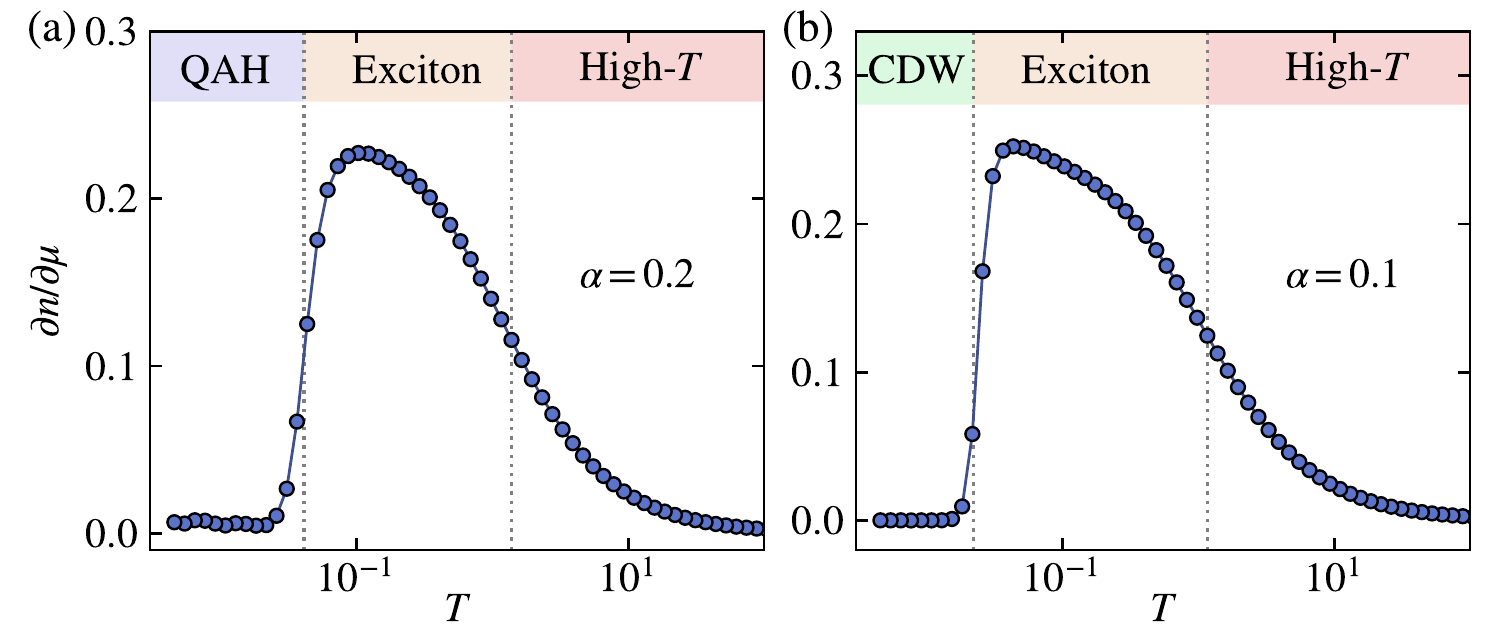}
\caption{The temperature-dependent electron compressibility $\partial n/\partial \mu$ of Kang-Vafek model in the cases of (a) $\alpha=0.2$, (b) $\alpha=0.1$. They are duplicates of Fig.3(c) and Fig.S5(c) to keep the present section self-consistent.}
\label{FigS:Compress}
\end{figure}

In this section, we add a chemical potential term to the original Kang-Vafek model
[Eq. (1) in the main text], i.e.
\begin{equation}
H = U_0\sum_{\varhexagon}({Q_{\varhexagon}} + \alpha T_{\varhexagon} - 1)^2 - \mu\sum_i \hat n_i,
\end{equation}
to account for the electronic response with chemical potential.
We calculate the averaged electron density
$$n_e (\mu, \beta) = \frac{N_e}{N} = \frac{1}{N}\frac{\mathrm{Tr}[\sum_i \hat n_i e^{-\beta H(\mu)}]}{\mathrm{Tr}[e^{-\beta H(\mu)}]},$$
for different inverse temperature $\beta=1/T$ and different chemical potentials $\mu$ ranging from
0 to 0.1.
As shown in \Fig{FigS:ParticleNum}, for both $\alpha=0.1$ and $\alpha=0.2$,
the particle density responds to chemical potential linearly above the transition temperature.
The electronic compressibility at $\mu=0$ is then approximately obtained via,
$$\frac{\partial n}{\partial \mu} \approx \frac{n_e(\Delta\mu,\beta)-n_e(0,\beta)}{\Delta\mu},$$
where $\Delta\mu=0.01$ is taken.
The obtained compressibility versus temperature is shown in \Fig{FigS:Compress} for two cases of
$\alpha=0.2$ and $\alpha=0.1$.
The results exhibit zero-value compressibility in both QAH and CDW low-$T$
phases due to their insulator nature, whereas show finite-value in the intermediate-$T$ regime
verifying the metallic nature.

\subsection{Section VII: Gaussian state approach to twisted bilayer graphene}

At finite temperature $T$, the imaginary time evolution equation \cite%
{shi2020} for density matrices $\rho $ reads%
\begin{equation}
d_{\tau }\rho =-\{F(\rho )-f(\rho ),\rho \},  \label{eq:f}
\end{equation}%
which guarantees the monotonic decrease of the free energy $f(\rho )=\mathrm{%
tr}(\rho F(\rho ))$\ with $F(\rho )=H+T\mathrm{ln}\rho $ being the free
energy operator.

We approximate the density matrix $\rho $ by the Gaussian state%
\begin{equation}
\rho _{\text{G}}=\frac{1}{Z}e^{-\frac{1}{2T}C^{\dagger }\Omega C},
\end{equation}%
where $Z=\mathrm{tr}(e^{-\frac{1}{2T}C^{\dagger }\Omega C})$ is the
partition function, $\Omega =\Omega ^{\dagger }$ is a matrix in the Nambu
basis $C^{\dagger }=(c_{1}^{\dagger },\ldots ,c_{N_{\text{f}}}^{\dagger
},c_{1},\ldots c_{N_{\text{f}}})$, the creation and annihilation operators $%
c_{i}^{\dagger }$ and $c_{i}$ fulfill the anti-commutation relation $%
\{c_{i},c_{j}^{\dagger }\}=\delta _{ij}$, and $N_{\text{f}}$ is the number
of fermionic modes. The density matrix $\rho _{\text{G}}$ is fully
characterized by its $2N_{\text{f}}\times 2N_{\text{f}}$ covariance matrix%
\begin{equation}
\Gamma =\mathrm{tr}(\rho _{\text{G}}CC^{\dagger })=\frac{1}{e^{-\frac{\Omega
}{T}}+1}.
\end{equation}%
By projecting Eq.~\eqref{eq:f} in the tangential space of the variational
manifold \cite{shi2018,shi2020}, we obtain EOM%
\begin{equation}
\partial _{\tau }\Gamma =\{\mathcal{F},\Gamma \}-2\Gamma \mathcal{F}\Gamma ,
\end{equation}%
where the mean-field free energy $\mathcal{F}=\mathcal{H}-\Omega $ is
determined by the mean-field Hamiltonian $\mathcal{H}_{ij}=-2\delta \mathcal{%
h}H\mathcal{i}_{\text{G}}/\delta \Gamma _{ij}$ and $\mathcal{h}H\mathcal{i}_{%
\text{G}}=tr(H\rho _{\text{G}})$.

For the Kang-Vafek (KV) model Eq.~({\color{blue}1}), the diagonal and off-diagonal
blocks%
\begin{align}
\varepsilon & =-M^{2}+2[1-\mathrm{tr}(\mathcal{h}cc^{\dagger }\mathcal{i}_{%
\text{G}}M)]M+2M\mathcal{h}cc^{\dagger }\mathcal{i}_{\text{G}}M,  \notag \\
\Delta & =-2M\mathcal{h}cc\mathcal{i}_{\text{G}}M,  \label{eq:meanH}
\end{align}%
of the mean-field Hamiltonian $\mathcal{H=}\left(
\begin{array}{cc}
\varepsilon & \Delta \\
\Delta ^{\dagger } & -\varepsilon ^{T}%
\end{array}%
\right) $ in the local basis $c_{i}$ of each hexagon are determined by%
\begin{equation}
M=\frac{1}{3}I_{6}+\left(
\begin{array}{cc}
0 & t_{ab} \\
t_{ab}^{\dagger } & 0%
\end{array}%
\right) ,t_{ab}=\alpha \left(
\begin{array}{ccc}
-1 & 0 & 1 \\
1 & -1 & 0 \\
0 & 1 & -1%
\end{array}%
\right) .
\end{equation}%
where the six sites in the hexagon are labeled in Fig.~\ref{hex}. The energy
per hexagon reads%
\begin{equation}
\mathcal{h}H_{\varhexagon}\mathcal{i}_{\text{G}}=[\mathrm{tr}((1-\mathcal{h}%
cc^{\dagger }\mathcal{i}_{\text{G}})M)-1]^{2}+\mathrm{tr}[(1-\mathcal{h}%
cc^{\dagger }\mathcal{i}_{\text{G}})M\mathcal{h}cc^{\dagger }\mathcal{i}_{%
\text{G}}M]+\mathrm{tr}[M\mathcal{h}c^{\dagger }c^{\dagger }\mathcal{i}_{%
\text{G}}M\mathcal{h}cc\mathcal{i}_{\text{G}}].  \label{eq:Hh}
\end{equation}%
\begin{figure}[tbph]
\centering \includegraphics[width=0.15\textwidth]{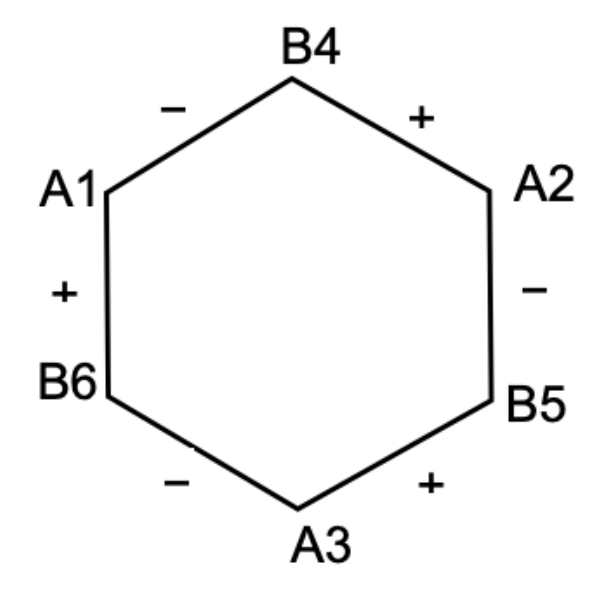}
\caption{The labels of the six sites in the hexagon.}
\label{hex}
\end{figure}

The free energy%
\begin{equation}
f(\rho )=\sum_{\varhexagon}\mathcal{h}H_{\varhexagon}\mathcal{i}_{\text{G}}-\frac{1}{2}\mathrm{%
tr}[(1-\Gamma )\Omega +T\mathrm{ln}(e^{-\frac{E}{T}}+1)]
\end{equation}%
decreases monotonically in the imaginary time evolution, where the diagonal
matrix $E=diag(E_{1},E_{2},...,E_{2N_{\text{f}}})$ is constructed by the
eigenvalues of $\Omega $. In the asymptotic limit, the density matrix
reaches the thermal equilibrium state, where $\mathcal{H}=\Omega $, and the
pairing term $\mathcal{h}c_{i}c_{j}\mathcal{i}_{\text{G}}=$ $\mathcal{h}%
c_{i}^{\dagger }c_{j}^{\dagger }\mathcal{i}_{\text{G}}=0$ due to the
repulsive Coulomb interaction inherited by the KV model. As a result, the
off-diagonal block $\Delta =0$.

In agreement with the XTRG approach, the finite-T phase diagram obtained by
the optimal mean-field theory consists of the QAH states, the CDW states,
and high-T trivial states. It is remarkable that the QAH and CDW states
possess a lot of symmetries which are represented by the structure of the
covariance matrix $\Gamma $. Therefore, the thermal state can be described
with very little order parameters, as we will show in the following.

Let us list several typical symmetries $S$ of the Hamiltonian. The
Hamiltonian has the particle hole symmetry $S_{c}$, the time reversal
symmetry $S_{T}$, the translational symmetry $P_{\overrightarrow{a}_{j}}$ $%
(j=1,2,3)$, the rotational symmetry $R_{3}$ with respect to the center of
the hexagon, the reflection symmetry $Z_{2h}$ with respect to the horizontal
axis $x_{1}$, the symmetry $n_{A}Z_{2v}$ of the reflection with respect to
the vertical axis $x_{2}$ and $c_{i}\rightarrow -c_{i}$ in one sub-lattice.
The vectors and axises in the honeycomb Moir\'{e} lattice are shown in Fig.~%
\ref{vectors}.
\begin{figure}[tbph]
\centering \includegraphics[width=0.5\textwidth]{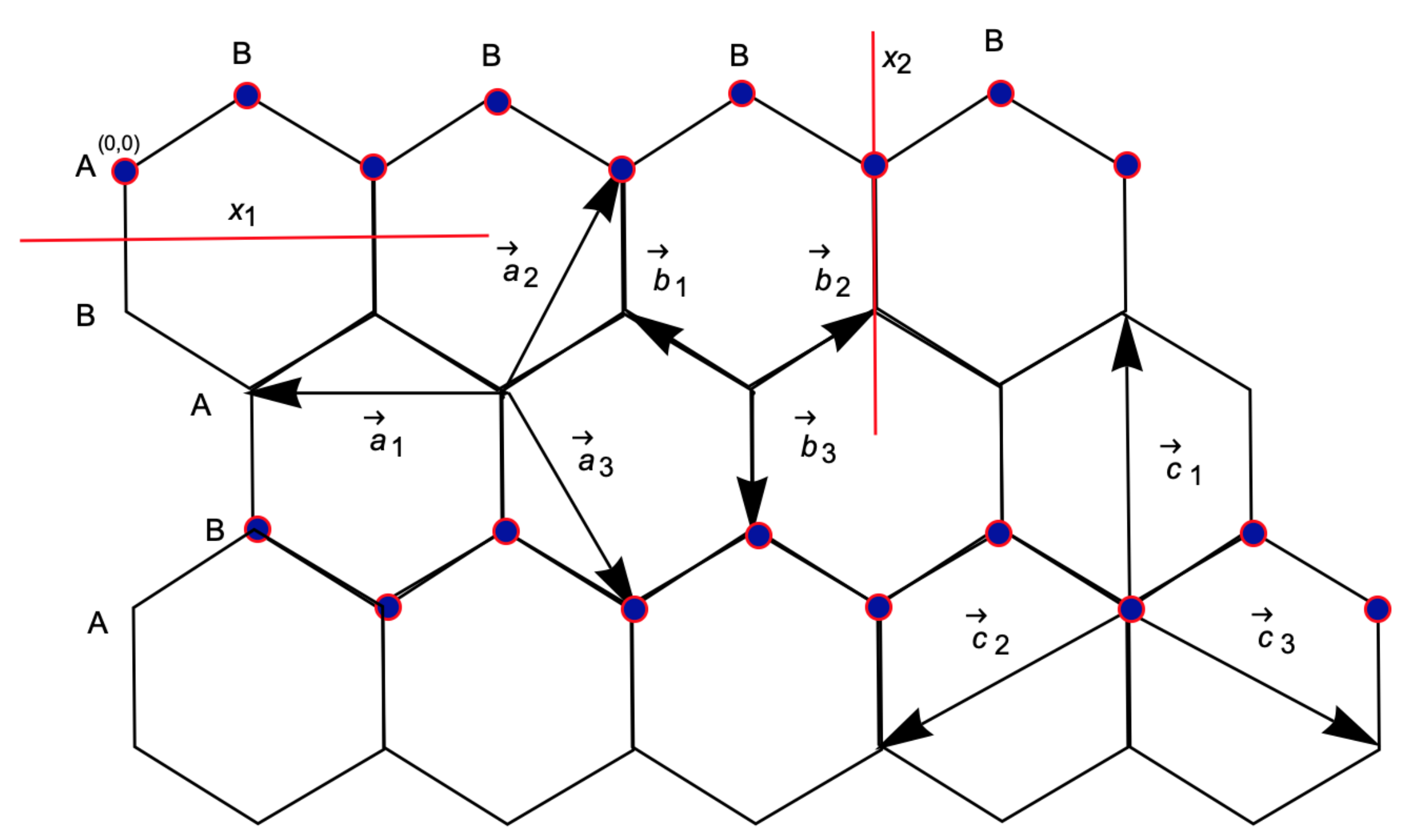}
\caption{The horizontal and vertical axis $x_{1}$ and $x_{2}$. The
sublattice $A$ of the honeycomb lattice are generated by linear combinations
of the basis vectors $\protect\overset{\rightarrow }{a}_{1}=(-\protect\sqrt{3%
},0),\text{ }\protect\overset{\rightarrow }{a}_{2}=\left( \frac{\protect%
\sqrt{3}}{2},\frac{3}{2}\right) $, $\protect\overset{\rightarrow }{A}\left(
n_{1}n_{2}\right) =n_{1}\protect\overset{\rightarrow }{a}_{1}+n_{2}%
\protect\overset{\rightarrow }{a}_{2}$, where $n_{1}$, $n_{2}$ are
intergers. The sublattices $A$ and $B$ are connected by the vectors $%
\protect\overset{\rightarrow }{b}_{1}=\left( \frac{-\protect\sqrt{3}}{2},%
\frac{1}{2}\right) ,\text{ }\protect\overset{\rightarrow }{b}_{2}=\left(
\frac{\protect\sqrt{3}}{2},\frac{1}{2}\right) ,\text{ }\protect\overset{%
\rightarrow }{b}_{3}=(0,-1)$. And here are other uesful vectors $%
\protect\overset{\rightarrow }{a}_{3}=\left( \frac{\protect\sqrt{3}}{2},-%
\frac{3}{2}\right) $, $\protect\overset{\rightarrow }{c}_{1}=(0,2),\text{ }%
\protect\overset{\rightarrow }{c}_{2}=\left( -\protect\sqrt{3},-1\right) ,%
\text{ }\protect\overset{\rightarrow }{c}_{3}=\left( \protect\sqrt{3}%
,-1\right) $.}
\label{vectors}
\end{figure}

The covariance matrix $\gamma =\mathcal{h}cc^{\dagger }\mathcal{i}_{\text{G}%
} $ in the coordinate space shows that the QAH state has symmetries $S_{c}$,
$P_{\overrightarrow{a}_{j}}(j=1,2,3)$, $R_{3}$ and the symmetry $S_{T}Z_{2h}$%
. The properties of $\gamma $ and the corresponding symmetry are listed as
follows:

\begin{itemize}
\item The particle-hole symmetry $S_{c}\Longrightarrow \gamma =1-\gamma ^{T}%
\overset{\gamma =\gamma ^{\dagger }}{\Longrightarrow }\gamma _{ii}=0.5,~ \mathrm{Re}(\gamma _{ij(i\neq j)})=0$.

\item The translational symmetry $P_{\overrightarrow{a}_{j}}$ $%
(j=1,2,3)\Longrightarrow $ The covariance matrices for all hexagons are
same, e.g., $\gamma _{41}=\gamma _{53}$.

\item The rotational symmetry $R_{3}\Longrightarrow$$\gamma_{41}=%
\gamma_{63}=\gamma_{52}$, $\gamma_{53}=\gamma_{42}=\gamma_{61}$, $%
\gamma_{21}=\gamma_{13}=\gamma_{32}$, $\gamma_{54}=\gamma_{46}=\gamma_{65}$,
$\gamma_{43}=\gamma_{62}=\gamma_{51}$.

\item The symmetry $S_{T}Z_{2h}\Longrightarrow \gamma _{23}=\gamma
_{54}^{\ast }$.
\end{itemize}

The symmetries play an important role since they completely determine the
covariance matrix%
\begin{equation}
\left\langle cc^{\dagger }\right\rangle _{\text{G}}=\gamma =\left(
\begin{array}{cc}
\gamma _{a} & \gamma _{\text{ab}} \\
\gamma _{\text{ab}}^{\dagger } & \gamma _{a}%
\end{array}%
\right) ,\gamma _{a}=\left(
\begin{array}{ccc}
0.5 & \delta ^{\ast } & \delta \\
\delta & 0.5 & \delta ^{\ast } \\
\delta ^{\ast } & \delta & 0.5%
\end{array}%
\right) ,\gamma _{\text{ab}}=\left(
\begin{array}{ccc}
\delta _{\text{ab}} & \delta _{\text{ab2}} & \delta _{\text{ab}} \\
\delta _{\text{ab}} & \delta _{\text{ab}} & \delta _{\text{ab2}} \\
\delta _{\text{ab2}} & \delta _{\text{ab}} & \delta _{\text{ab}}%
\end{array}%
\right)
\end{equation}%
with only three independent order parameters%
\begin{align}
\delta _{ab}& =\mathcal{h}b_{\overrightarrow{A}+\overrightarrow{b}%
_{j}}^{\dagger }a_{\overrightarrow{A}}\mathcal{i}_{\text{G}},  \notag \\
\delta & =\mathcal{h}a_{\overrightarrow{A}+\overrightarrow{a}_{j}}^{\dagger
}a_{\overrightarrow{A}}\mathcal{i}_{\text{G}}=\mathcal{h}b_{\overrightarrow{A}+%
\overrightarrow{b}_{1}+\overrightarrow{a}_{j}}b_{\overrightarrow{A}+%
\overrightarrow{b}_{1}}^{\dagger }\mathcal{i}_{\text{G}},  \notag \\
\delta _{ab2}& =\mathcal{h}b_{\overrightarrow{A}+\overrightarrow{c}%
_{j}}^{\dagger }a_{\overrightarrow{A}}\mathcal{i}_{\text{G}},\quad j=1,2,3,
\label{Eq:qahOr}
\end{align}%
where $a_{\overrightarrow{r_{a}}}$ and $b_{\overrightarrow{r_{b}}}$ are the
annihilation operators of electrons in sublattices $A$ and $B$.

The covariance matrix gives rise to the mean-field Hamiltonian%
\begin{equation}
\varepsilon =\left(
\begin{array}{cc}
T_{a} & T_{\text{ab}} \\
T_{\text{ab}}^{\dagger } & T_{a}%
\end{array}%
\right) ,T_{a}=\left(
\begin{array}{ccc}
0 & t_{2} & t_{2}^{\ast } \\
t_{2}^{\ast } & 0 & t_{2} \\
t_{2} & t_{2}^{\ast } & 0%
\end{array}%
\right) ,T_{ab}=\left(
\begin{array}{ccc}
\bar{t}_{1}^{\ast } & t_{3}^{\ast } & \bar{t}_{1}^{\ast } \\
\bar{t}_{1}^{\ast } & \bar{t}_{1}^{\ast } & t_{3}^{\ast } \\
t_{3}^{\ast } & \bar{t}_{1}^{\ast } & \bar{t}_{1}^{\ast }%
\end{array}%
\right)
\end{equation}
in each hexagon through Eq.~\eqref{eq:meanH}, with the effective hopping
strengths%
\begin{align}
\bar{t}_{1}& =2\left( \alpha ^{2}+\frac{1}{9}\right) \delta _{ab}-\frac{4}{3}%
\alpha \delta -2\alpha ^{2}\delta _{ab2},  \notag \\
t_{2}& =2\left( -\frac{2\alpha \delta _{ab}}{3}+\left( 3\alpha ^{2}+\frac{1}{%
9}\right) \delta +\frac{2\alpha \delta _{ab2}}{3}\right) ,  \notag \\
t_{3}& =2\left( -2\alpha ^{2}\delta _{ab}+\frac{4\alpha \delta }{3}+\left(
2\alpha ^{2}+\frac{1}{9}\right) \delta _{ab2}\right) .
\end{align}%
between the nearest neighbor, the second-neighbor, and the third-neighbor
sites, as shown in Fig.~\ref{int}.
\begin{figure}[tbph]
\centering \includegraphics[width=0.8\textwidth]{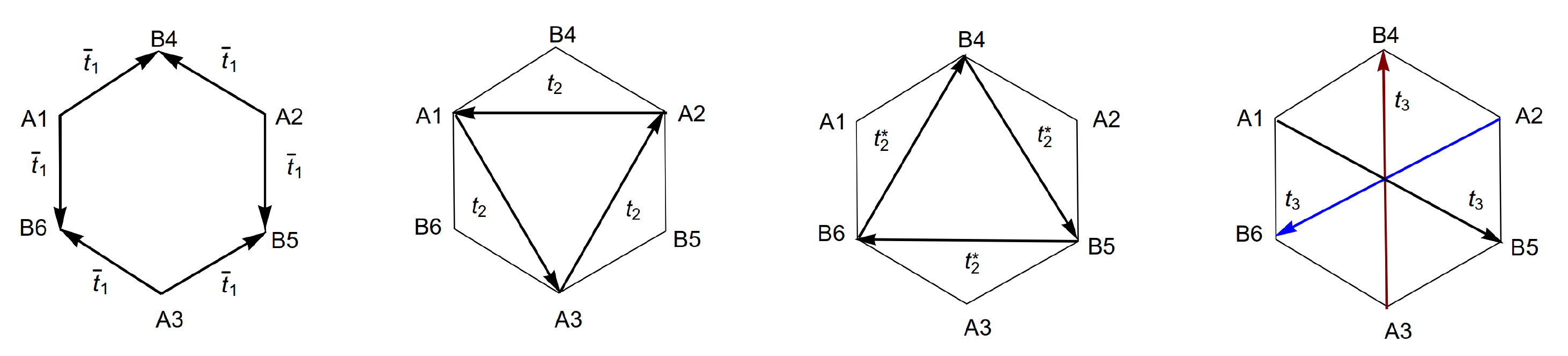}
\caption{The effective hopping strengths of the mean field Hamiltonian in
QAH phases.}
\label{int}
\end{figure}

In the second quantized form, the mean-field Hamiltonian of the honeycomb
Moir\'{e} lattice reads%
\begin{equation}
H=\sum_{\overset{\rightarrow }{A}}\sum_{j}\left[ t_{1}b_{\overset{%
\rightarrow }{A}+\overset{\rightarrow }{b}_{j}}^{\dagger }a_{\overset{%
\rightarrow }{A}}+t_{2}a_{\overset{\rightarrow }{A}+\overset{\rightarrow }{a}%
_{j}}^{\dagger }a_{\overset{\rightarrow }{A}}+t_{2}^{\ast }b_{\overset{%
\rightarrow }{A}+\overset{\rightarrow }{b_{1}}+\overset{\rightarrow }{a}%
_{j}}^{\dagger }b_{\overset{\rightarrow }{A}+\overset{\rightarrow }{b_{1}}%
}+t_{3}b_{\overset{\rightarrow }{A}+\overset{\rightarrow }{c}_{j}}^{\dagger
}a_{\overset{\rightarrow }{A}}+h.c.\right] ,  \label{RealH}
\end{equation}%
where $t_{1}=2\bar{t}_{1}$. The Fourier transforms $a_{k}=\sum_{\overset{%
\rightarrow }{A}}e^{-i\overset{\rightarrow }{k}\cdot \overset{\rightarrow }{A%
}}a_{\overset{\rightarrow }{A}}/\sqrt{N}$ and $b_{k}=\sum_{\overset{%
\rightarrow }{A}}e^{-i\overset{\rightarrow }{k}\cdot \left( \overset{%
\rightarrow }{A}+\overset{\rightarrow }{b}_{j}\right) }b_{\overset{%
\rightarrow }{A}+\overset{\rightarrow }{b}_{j}}/\sqrt{N}$ result in the
mean-field Hamiltonian $H=\sum_{\overset{\rightarrow }{k}}c_{k}^{\dagger }h^{%
\text{QAH}}(k)c_{k}$ in the momentum space $c_{k}=\left( a_{k},b_{k}\right)
^{T}$, where%
\begin{equation}
h^{\text{QAH}}(k)=\left(
\begin{array}{cc}
h_{k,11} & h_{k,12} \\
h_{k,12}^{\ast } & -h_{k,11}%
\end{array}%
\right)
\end{equation}%
is determined by%
\begin{eqnarray}
h_{k,11} &=&t_{2}\sum_{j}e^{-i\overset{\rightarrow }{k}\cdot \overset{%
\rightarrow }{a_{j}}}+t_{2}^{\ast }\sum_{j}e^{i\overset{\rightarrow }{k}%
\cdot \overset{\rightarrow }{a_{j}}},  \notag \\
h_{k,12} &=&t_{1}^{\ast }\sum_{j}e^{i\overset{\rightarrow }{k}\cdot \overset{%
\rightarrow }{b_{j}}}+t_{3}^{\ast }\sum_{j}e^{i\overset{\rightarrow }{k}%
\cdot \overset{\rightarrow }{c_{j}}}.
\end{eqnarray}%
The covariance matrix%
\begin{equation}
\left\langle c_{k}c_{k}^{\dagger }\right\rangle _{\text{G}}=\frac{1}{e^{-\frac{1}{T} h^{%
\text{QAH}}(k)}+1}=\frac{h^{\text{QAH}}(k)}{2 d_{k,c}}\left(
1-2f_{k}^{c}\right) +\frac{1}{2}I_{2}
\end{equation}%
can be obtained by the diagonalization $U_{k}^{\dagger }h^{\text{QAH}%
}(k)U_{k}=diag(d_{k,c},d_{k,v})$ of the mean-field Hamiltonian $h^{\text{QAH}%
}(k)$ using the unitary transformation $U_{k}$, where the Fermi-Dirac
distribution $f_{k}^{\beta }=1/(e^{d_{k,\beta }/T}+1)$ is determined by the
dispersion relations $d_{k,(c,v)}=\pm \sqrt{h_{k,11}^{2}+\lvert
h_{k,12}\rvert ^{2}}$ in the conduction ($v$) and valance ($v$) bands at the
temperature $T$.

The covariance matrix gives rise to the self-consistent equations%
\begin{align}
\delta _{ab}& =\frac{1}{2N}\sum_{k}e^{-i\overset{\rightarrow }{k}\cdot
\overset{\rightarrow }{b}_{j}}\frac{-h_{12}}{d_{k,c}}\tanh \left( \frac{d_{k,c}}{2T%
}\right) ,  \notag \\
\delta & =\frac{1}{2N}\sum_{k}e^{-i\overset{\rightarrow }{k}\cdot \overset{%
\rightarrow }{a}_{j}}\frac{-h_{11}}{d_{k,c}}\tanh \left( \frac{d_{k,c}}{2T}%
\right) ,  \notag \\
\delta _{ab2}& =\frac{1}{2N}\sum_{k}e^{-i\overset{\rightarrow }{k}\cdot
\overset{\rightarrow }{c}_{j}}\frac{-h_{12}}{d_{k,c}}\tanh \left( \frac{d_{k,c}}{2T%
}\right)  \label{eq:DT}
\end{align}%
for the order parameters. Close to the phase transition, all the order
parameters tend to zero, and Eq.~\eqref{eq:DT} can be linearized as%
\begin{equation}
\left(
\begin{array}{c}
\delta _{\text{ab}} \\
\delta \\
\delta _{\text{ab2}}%
\end{array}%
\right) =\frac{1}{4T}\left(
\begin{array}{c}
t_{1} \\
t_{2} \\
t_{3}%
\end{array}%
\right) =\frac{1}{T}M_{\text{QAH}}\left(
\begin{array}{c}
\delta _{\text{ab}} \\
\delta \\
\delta _{\text{ab2}}%
\end{array}%
\right) .
\end{equation}%
The critical temperature $T_{c}$ is thus determined by the largest positive
eigenvalue $\lambda _{\mathrm{max}}$ of%
\begin{equation*}
M_{\text{QAH}}=\frac{1}{2}\left(
\begin{array}{ccc}
2\alpha ^{2}+\frac{2}{9} & -\frac{4}{3}\alpha & -2\alpha ^{2} \\
-\frac{2}{3}\alpha & 3\alpha ^{2}+\frac{1}{9} & \frac{2}{3}\alpha \\
-2\alpha ^{2} & \frac{4}{3}\alpha & 2\alpha ^{2}+\frac{1}{9}%
\end{array}%
\right) .
\end{equation*}

Furthermore, using symmetries we can significantly reduce the degrees of
freedom in the $N_{\mathrm{f}}\times N_{\mathrm{f}}$ covariance matrix to
three order parameters, as a result, the above analysis can be applied to
the system in the thermodynamic limit. To speed up the calculation, we first
numerically the flow equations of $\gamma $ for a small system, and achieve
the three order parameters in the thermal state. Using the order parameters
of the small system as the initial condition, we solve the non-linear Eq.~%
\eqref{eq:DT} to obtain the order parameters for the system in the
thermodynamic limit, where the free energy density%
\begin{equation}
f(\delta _{\text{ab}},\delta ,\delta _{\text{ab2}})=\mathcal{h}H_{\varhexagon%
}\mathcal{i}_{\text{G}}+\frac{1}{N_{f}}\sum_{k}2T\left[ f_{k}^{c}\mathrm{ln}%
f_{k}^{c}+f_{k}^{v}\mathrm{ln}f_{k}^{v}\right]
\end{equation}%
is minimized.

In the CDW phase, the density distribution of the CDW state has three
equivalent spatial structures for the system in the thermodynamic limit,
which possesses the symmetries $P_{\overrightarrow{a}_{1}}$, $P_{%
\overrightarrow{a}_{2}}$ and $P_{\overrightarrow{a}_{3}}$, respectively. We
analyze one of the three equivalent CDW states that maintains the symmetry $%
P_{\overrightarrow{a}_{1}}$ without loss of generality. The structure of the covariance matrices $\gamma _{\mathrm{odd}}\equiv\lambda$
and $\gamma _{\mathrm{even}}$ in the odd and even rows, respectively, indicates that the CDW state possesses the
symmetries $S_{T}$, $P_{\overrightarrow{a}_{1}}$, $n_{A}Z_{2v}$, and the
combination symmetries $S_{c}P_{\overrightarrow{a}_{3}}$, $S_{c}Z_{2h}$. The
properties of the covariance matrices and the corresponding symmetry are listed as follows.

\begin{itemize}
\item The time reversal symmetry $S_{T}\Longrightarrow \lambda =\lambda
^{\ast }\overset{\lambda =\lambda ^{\dagger }}{\Longrightarrow }$ $\gamma _{%
\mathrm{odd}}$ and $\gamma _{\mathrm{even}}$ are real symmetric matrices.

\item The translational symmetry $P_{\overrightarrow{a}_{1}}\Longrightarrow $
The covariance matrices for each rows are same, e.g., $\lambda _{11}=\lambda _{22}$.

\item The combination symmetry $S_{c}P_{\overrightarrow{a}%
_{3}}\Longrightarrow \gamma _{\mathrm{odd}}=I-\gamma _{\mathrm{even}}^{T}$, $%
\lambda _{14}=-\lambda _{35}$, $\lambda _{44}=1-\lambda_{55}$.

\item The combination symmetry $S_{c}Z_{2h}\Longrightarrow \lambda
_{11}=1-\lambda _{66}$, $\lambda _{22}=1-\lambda _{55}$, $\lambda _{44}=1-\lambda
_{33}$; $\lambda _{16}=\lambda _{25}=0$, $\lambda _{14}=-\lambda _{63}$, $\lambda
_{24}=-\lambda _{53}$; $\lambda _{12}=-\lambda _{56}$, $\lambda _{13}=-\lambda
_{64}$, $\lambda _{23}=-\lambda _{54}$; $\lambda _{15}=-\lambda _{62}$, $\lambda
_{34}=-\lambda _{43}=0$.

\item The symmetry $n_{A}Z_{2v}\Longrightarrow \lambda _{13}=\lambda _{23}$, $%
\lambda_{36}=-\lambda _{35}$.
\end{itemize}

Therefore, the covariance matrix $\gamma $ only has four independent order
parameters%
\begin{align}
n_{0}& =\mathcal{h}a_{\overrightarrow{A_{o}}}a_{\overrightarrow{A_{o}}%
}^{\dagger }\mathcal{i}_{\text{G}}-\frac{1}{2},  \notag \\
n_{1}& =\mathcal{h}a_{\overrightarrow{A_{o}}}a_{\overrightarrow{A_{o}}-%
\overrightarrow{a}_{1}}^{\dagger }\mathcal{i}_{\text{G}},  \notag \\
n_{2}& =\mathcal{h}a_{\overrightarrow{A_{o}}}a_{\overrightarrow{A_{o}}+%
\overrightarrow{a}_{3}}^{\dagger }\mathcal{i}_{\text{G}},  \notag \\
\Delta _{0}& =\mathcal{h}a_{\overrightarrow{A_{o}}}b_{\overrightarrow{A_{o}}+%
\overrightarrow{c}_{2}}^{\dagger }\mathcal{i}_{\text{G}},  \label{Eq:cdwOr}
\end{align}%
where $\overrightarrow{A_{o}}=l_{1}\overrightarrow{a}_{1}+2l_{2}%
\overrightarrow{a}_{3}$ and $l_{1,2}$ are integers. The covariance matrix%
\begin{equation}
\gamma _{\text{odd}}=\left(
\begin{array}{cccccc}
n_{0}+\frac{1}{2} & n_{1} & n_{2} & 0 & -\Delta _{0} & 0 \\
n_{1} & n_{0}+\frac{1}{2} & n_{2} & 0 & 0 & \Delta _{0} \\
n_{2} & n_{2} & \frac{1}{2}-n_{0} & 0 & 0 & 0 \\
0 & 0 & 0 & n_{0}+\frac{1}{2} & -n_{2} & -n_{2} \\
-\Delta _{0} & 0 & 0 & -n_{2} & \frac{1}{2}-n_{0} & -n_{1} \\
0 & \Delta _{0} & 0 & -n_{2} & -n_{1} & \frac{1}{2}-n_{0}%
\end{array}%
\right)
\end{equation}%
results in the mean-field Hamiltonian%
\begin{equation}
\varepsilon ^{o}=\left(
\begin{array}{cccccc}
e_{1} & e_{3} & e_{4} & e_{5} & e_{6} & 0 \\
e_{3} & e_{1} & e_{4} & -e_{5} & 0 & -e_{6} \\
e_{4} & e_{4} & e_{2} & 0 & e_{5} & -e_{5} \\
e_{5} & -e_{5} & 0 & -e_{2} & -e_{4} & -e_{4} \\
e_{6} & 0 & e_{5} & -e_{4} & -e_{1} & -e_{3} \\
0 & -e_{6} & -e_{5} & -e_{4} & -e_{3} & -e_{1}%
\end{array}%
\right)
\end{equation}
in odd rows with the hopping strengths%
\begin{align}
e_{1}& =\frac{2}{9}n_{0}+4\alpha ^{2}n_{2},  \notag \\
e_{2}& =-\frac{2}{9}n_{0}+2\alpha ^{2}\left(
-2n_{0}+2n_{1}\right) ,  \notag \\
e_{3}& =\frac{2}{9}n_{1}+\frac{4\alpha }{3}\Delta _{0}+\alpha ^{2}\left(
-2n_{0}+2n_{1}-4n_{2}\right) ,  \notag \\
e_{4}& =\frac{2}{9}n_{2}-\frac{2\alpha }{3}\Delta _{0}-\alpha ^{2}\left(
-2n_{0}+2n_{1}\right) ,  \notag \\
e_{5}& =-\frac{2}{3}\alpha \left( 2n_{0}-n_{1}+n_{2}-3\alpha \Delta
_{0}\right) ,  \notag \\
e_{6}& =\frac{2}{9}\left( 6\alpha \left( n_{2}-n_{1}\right) -\left(
1+9\alpha ^{2}\right) \Delta _{0}\right) .
\end{align}%
The combination symmetry $S_{c}P_{\overrightarrow{a}_{3}}$ determines the
relation between the mean-field Hamiltonian in odd and even rows as $%
\varepsilon ^{o}=-\varepsilon ^{e}$.

In the second quantized form, the mean-field Hamiltonian of the honeycomb
Moir\'{e} lattice reads%
\begin{equation}
H=\sum_{\varhexagon^{o}}\sum_{i,j\in \varhexagon^{o}}c_{i}^{\dagger }\varepsilon
_{i,j}^{o}c_{j}+\sum_{\varhexagon^{e}}\sum_{i,j\in \varhexagon^{e}}c_{i}^{\dagger }\varepsilon
_{i,j}^{e}c_{j}.
\end{equation}%
The stripe has the translational symmetry $P_{\overrightarrow{a_{1}}}$ and $%
P_{2\overrightarrow{a_{3}}}$, thus, particles with the momentum $k_{h}\in
\{[0,2\pi _{x}/\sqrt{3}],[0,2\pi _{y}/3]\}$ hybridize with particles with
the momentum $k_{h_{\pm }}=k_{h}\pm 2\pi _{y}/3$ and the Brillouin zone
shrinks to $k_{h}\in \{[0,2\pi _{x}/\sqrt{3}],[0,2\pi _{y}/3]\}$, which is
different from the situation in QAH phases. The Fourier transformation gives
rise to the mean-field Hamiltonian $H=\sum_{k_{h}}c_{k_{h}}^{\dagger }h^{%
\text{CDW}}\left( k_{h}\right) c_{k_{h}}$ in the momentum space $%
c_{k_{h}}=(a_{k_{h}},b_{k_{h}},a_{k_{h_{+}}},b_{k_{h_{+}}})$, where%
\begin{equation}
h^{\text{CDW}}\left( k_{h}\right) =\left(
\begin{array}{cc}
0 & \bar{h}_{0}^{\dagger }\left( k_{h}\right) \\
\bar{h}_{0}\left( k_{h}\right) & 0%
\end{array}%
\right)
\end{equation}%
is determined by%
\begin{equation}
\bar{h}_{0}\left( k_{h}\right) =\left(
\begin{array}{cc}
\varepsilon _{k_{h}} & \bar{\varepsilon}_{k_{h}} \\
-e^{-i\frac{\pi }{3}}\bar{\varepsilon}_{k_{h}}^{\ast } & e^{-i\frac{\pi }{3}%
}\varepsilon _{k_{h}}^{\ast }%
\end{array}%
\right) ,
\end{equation}%
and%
\begin{align}
\varepsilon _{k}& =2e_{1}-e_{2}+2e_{3}\cos (\sqrt{3}k_{x})-4ie_{4}\cos (%
\frac{\sqrt{3}}{2}k_{x})\sin (\frac{3}{2}k_{y}),  \notag \\
\bar{\varepsilon}_{k}& =2ie^{-ik_{y}}e_{6}\sin (\sqrt{3}k_{x}).
\end{align}

The covariance matrix%
\begin{equation}
\left\langle \left(
\begin{array}{c}
c_{k} \\
c_{k^{\prime }}%
\end{array}%
\right) \left(
\begin{array}{cc}
c_{k}^{\dagger } & c_{k^{\prime }}^{\dagger }%
\end{array}%
\right) \right\rangle _{\mathrm{G}}=\frac{1}{e^{-\frac{1}{T} h^{\text{CDW}}(k)}+1}=\left(
\begin{array}{cc}
\frac{1}{2} & \frac{1}{2E_{k}}\bar{h}_{0}^{\dagger }(k)\left( 1-2f_{k}\right)
\\
\frac{1}{2E_{k}}\bar{h}_{0}(k)\left( 1-2f_{k}\right) & \frac{1}{2}%
\end{array}%
\right)
\end{equation}%
can be obtained by the diagonalization $U_{k}^{\dagger }h^{\text{CDW}%
}(k)U_{k}=diag(E_{k},E_{k},-E_{k},-E_{k})$ of the mean-field Hamiltonian $h^{%
\text{CDW}}(k)$ using the unitary transformation $U_{k}$, where the
Fermi-Dirac distribution in the CDW phase $f_{k}=1/(e^{E_{k}/T}+1)$ is
determined by the dispersion relation $E_{k}=\sqrt{\lvert \varepsilon
_{k}\rvert ^{2}+\lvert \bar{\varepsilon}_{k}\rvert ^{2}}$ at the temperature
$T$.

The covariance matrix leads to the self-consistent equations%
\begin{align}
n_{0}& =\frac{1}{N}\sum_{k}\frac{\varepsilon _{k}}{2E_{k}}\tanh
\left( \frac{E_{k}}{2T}\right) ,  \notag \\
n_{1}& =\frac{1}{N}\sum_{k}e^{ik\overset{\rightarrow }{a}_{1}}\frac{%
\varepsilon _{k}}{2E_{k}}\tanh \left( \frac{E_{k}}{2T}\right) ,  \notag \\
n_{2}& =\frac{1}{N}\sum_{k}e^{-ik\overset{\rightarrow }{a}_{3}}\frac{%
\varepsilon _{k}}{2E_{k}}\tanh \left( \frac{E_{k}}{2T}\right) ,  \notag \\
\delta _{0}& =\frac{1}{N}\sum_{k}e^{-ik\overset{\rightarrow }{c}_{3}}\frac{-%
\bar{\varepsilon}_{k}}{2E_{k}}\tanh \left( \frac{E_{k}}{2T}\right)
\label{eq:DT2}
\end{align}%
for the order parameters. Close to the phasae transition, all the order
parameters tend to zero, and Eq.~\eqref{eq:DT2} can be linearized as%
\begin{align}
4T n_{0}& =2e_{1}-e_{2}=\left( \frac{2}{3}+4\alpha
^{2}\right)  n_{0} -4\alpha ^{2}n_{1}+8\alpha
^{2}n_{2},  \notag \\
4Tn_{1}& =e_{3}=-2\alpha ^{2} n_{0}+\left( \frac{2%
}{9}+2\alpha ^{2}\right) n_{1}-4\alpha ^{2}n_{2}-\frac{4\alpha }{3}\delta
_{0},  \notag \\
4Tn_{2}& =e_{4}=2\alpha ^{2} n_{0} -2\alpha
^{2}n_{1}+\frac{2}{9}n_{2}+\frac{2\alpha }{3}\delta _{0},  \notag \\
4T\delta _{0}& =e_{6}=\frac{2}{9}[-6\alpha n_{1}+6\alpha n_{2}+\left(
1+9\alpha ^{2}\right) \delta _{0}].
\end{align}%
The critical temperature $T_{c}=\lambda _{\max }/4$ is determined by the
largest positive eigenvalue $\lambda _{\max }$ of the matrix%
\begin{equation}
M_{\mathrm{CDW}}=\left(
\begin{array}{cccc}
\frac{2}{3}+4\alpha ^{2} & -4\alpha ^{2} & 8\alpha ^{2} & 0 \\
-2\alpha ^{2} & \frac{2}{9}+2\alpha ^{2} & -4\alpha ^{2} & \frac{-4}{3}\alpha
\\
2\alpha ^{2} & -2\alpha ^{2} & \frac{2}{9} & \frac{2}{3}\alpha \\
0 & \frac{-4}{3}\alpha & \frac{4}{3}\alpha & \frac{2}{9}+2\alpha ^{2}%
\end{array}%
\right) .
\end{equation}

Finally, the finite-T phase diagram~\ref{phase} under the self-consistent
mean-field approximation can be obtained, where the first-order transition
between the QAH and CDW phases is displayed.
\begin{figure}[tbph]
\centering \includegraphics[width=0.4\textwidth]{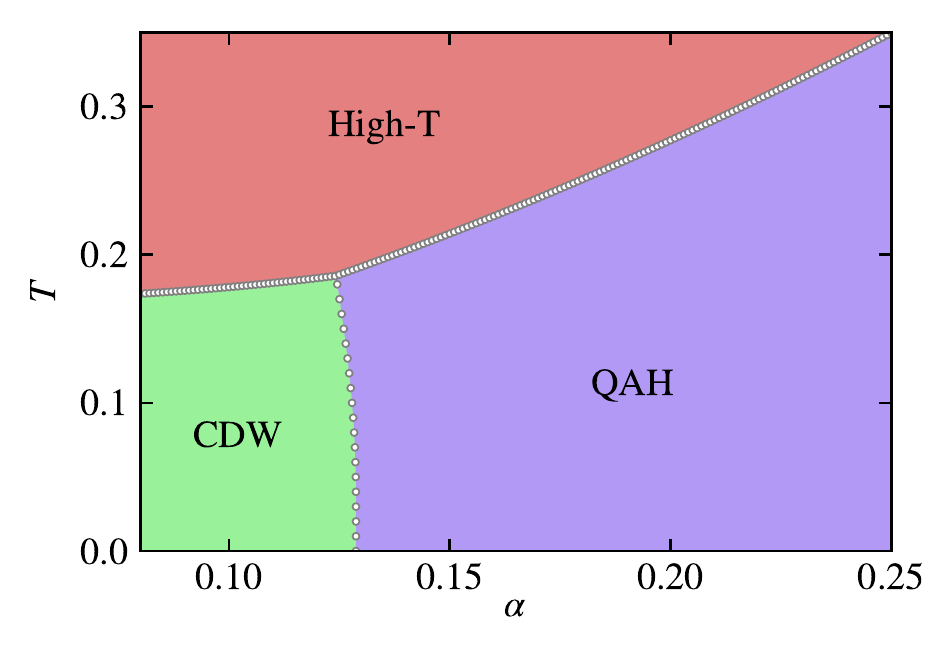}
\caption{The finite-T phase diagram obtained by the Gaussian state theory.}
\label{phase}
\end{figure}

\subsection{Section VIII: Exciton spectra and self-energy corrections}

At zero temperature, the single-particle Green's function from the Gaussian
state approach (the self-consistent mean-field theory) agrees excellently
with the result from DMRG~\cite{Chen2020TMI}. With respect to the Gaussian
thermal state $\rho _{G}$ described by three order parameters in QAH phases,
the KV Hamiltonian~\eqref{Eq:normor} is decomposed into normal ordered terms
via the Wick theorem as%
\begin{align}
H& =\mathcal{h}H\mathcal{i}+\sum_{\varhexagon}\text{:}c_{i}^{\dagger
}M_{ij}c_{j}\text{:}+\sum_{\varhexagon}\text{:}c_{i}^{\dagger
}M_{ij}c_{j}c_{l}^{\dagger }M_{lm}c_{m}\text{:},  \notag \\
& =\mathcal{h}H\mathcal{i}+\sum_{k}\text{:}c_{k,\alpha }^{\dagger }h_{\alpha
\beta }^{\mathrm{QAH}}(k)c_{k,\beta }\text{:}+\frac{1}{N}%
\sum_{k,p,q}(V_{k-q}^{\dagger }MV_{k})_{\alpha \beta }(V_{p}^{\dagger
}MV_{p-q})_{\overline{\alpha }\overline{\beta }}\text{:}c_{k-q,\alpha
}^{\dagger }c_{k,\beta }c_{p,\overline{\alpha }}^{\dagger }c_{p-q,\overline{%
\beta }}\text{:},  \label{Eq:normor}
\end{align}%
where the Fourier transformation $V_{k}=\left(\begin{array}{cc}e^{ikr_{A}^{h}} & 0\\0 & e^{ikr_{B}^{h}}\end{array}\right)$ is determined by the relative distance $r_{A}^{h}=\left((0,0),(\sqrt{3},0),(\sqrt{3}/2,-3/2)\right)^{T}$ and $r_{B}^{h}=\left((\sqrt{3}/2,1/2),(\sqrt{3},-1),(0,-1)\right)^{T}$, and we use the
Einstein summation convention for the Greek alphabet.

\begin{figure}[htbp]
\centering \includegraphics[width=0.8\textwidth]{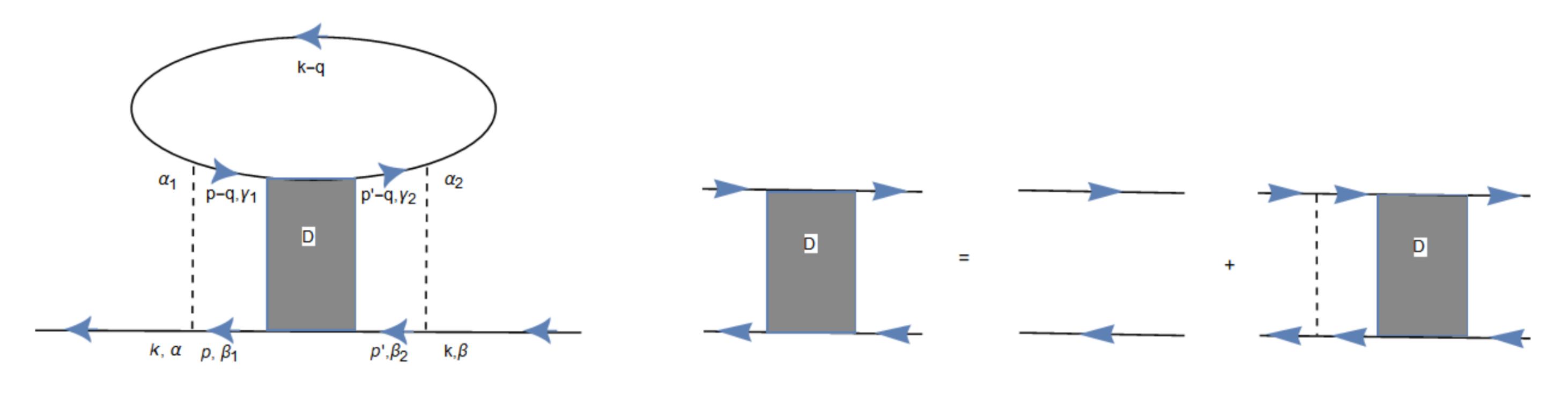}
\caption{The Hartree-like self energy $\Sigma_{k,\protect\alpha\protect\beta}^{(\text{H})} $.}
\label{Gqah}
\end{figure}

At finite temperature, the two-particle spectrum can be obtained by the ladder
diagram approximation beyond the mean-field theory, as shown in Fig.~\ref%
{Gqah}, and the self-energy correction, e.g., the Hartree
term, becomes%
\begin{align}
\Sigma _{k,\alpha \beta }^{(\text{H})}& =-\left( \frac{2}{N}\right)
^{2}\sum_{pp^{\prime }q}T\sum_{\omega _{q}}G_{k-q}^{(0)\alpha _{1}\alpha
_{2}}(\omega _{k}-\omega _{q})(V_{k}^{\dagger }MV_{p})_{\alpha \beta
_{1}}(V_{p-q}^{\dagger }MV_{k-q})_{\alpha _{1}\gamma _{1}}^{T}  \notag \\
& D_{pp^{\prime }}^{\beta _{1}\gamma _{1},\beta _{2}\gamma _{2}}(q,\omega
_{q})(V_{k-q}^{\dagger }MV_{p^{\prime }-q})_{\gamma _{2}\alpha
_{2}}^{T}(V_{p^{\prime }}^{\dagger }MV_{k})_{\beta _{2}\beta },  \label{se}
\end{align}%
where the free single-particle Green function $G_{k}^{(0)}(\omega
_{k})=1/(i\omega _{k}-h^{\mathrm{QAH}}(k))$. The exciton spectrum can be
determined by the density-density correlation function%
\begin{align}
D_{pp^{\prime }}^{\beta _{1}\gamma _{1},\beta _{2}\gamma _{2}}(q,\omega
_{q})& =\int_{0}^{\beta }d\tau D_{pp^{\prime }}^{\beta _{1}\gamma _{1},\beta
_{2}\gamma _{2}}(q,\tau )e^{i\omega _{q}\tau },  \notag \\
D_{pp^{\prime }}^{\beta _{1}\gamma _{1},\beta _{2}\gamma _{2}}(q,\tau )& =-%
\mathcal{h}T_{\tau }c_{p-q,\gamma _{1}}^{\dagger }(\tau )c_{p,\beta
_{1}}(\tau )c_{p^{\prime },\beta _{2}}^{\dagger }(0)c_{p^{\prime }-q,\gamma
_{2}}(0)\mathcal{i}.
\end{align}

It follows from the Heisenberg EOM that%
\begin{equation}
-\partial _{\tau }D_{pp^{\prime }}^{\beta _{1}\gamma _{1},\beta _{2}\gamma
_{2}}(q,\tau )=(\mathcal{h}c_{p-q,\gamma _{1}}^{\dagger }c_{p-q,\gamma _{2}}%
\mathcal{i}_{\text{G}}\delta _{\beta _{1},\beta _{2}}-\mathcal{h}c_{p,\beta
_{2}}^{\dagger }c_{p,\beta _{1}}\mathcal{i}_{\text{G}}\delta _{\gamma _{1},\gamma
_{2}})\delta _{p,p^{\prime }}\delta (\tau )+\sum_{k^{\prime}}\mathcal{M}_{p,k^{\prime
}}^{\beta _{1}\gamma _{1},\beta \gamma }(q)D_{k^{\prime }p^{\prime }}^{\beta
\gamma ,\beta _{2}\gamma _{2}}(q,\tau ),  \label{HD}
\end{equation}%
where%
\begin{align}
\mathcal{M}_{p,k^{\prime }}^{\beta _{1}\gamma _{1},\beta \gamma }(q)&
=(h_{\beta _{1}\beta }^{\mathrm{QAH}}(p)\delta _{\gamma _{1}\gamma }-\delta _{\beta
_{1}\beta }h_{\gamma \gamma _{1}}^{\mathrm{QAH}}(p-q))\delta _{p,k^{\prime }}  \notag \\
& +\frac{2}{N}[\mathcal{h}c_{p,\overline{\beta }}^{\dagger }c_{p,\beta _{1}}%
\mathcal{i}_{\text{G}}(V_{p}^{\dagger }MV_{k^{\prime }})_{\overline{\beta }\beta
}(V_{k^{\prime }-q}^{\dagger }MV_{p-q})_{\gamma \gamma _{1}}  \notag \\
& -(V_{p}^{\dagger }MV_{k^{\prime }})_{\beta _{1}\beta }\mathcal{h}%
c_{p-q,\gamma _{1}}^{\dagger }c_{p-q,\overline{\gamma }}\mathcal{i}%
_{\text{G}}(V_{k^{\prime }-q}^{\dagger }MV_{p-q})_{\gamma \overline{\gamma }}].
\end{align}

We define the correlation function $\bar{D}_{pp^{\prime }}^{\beta
_{1}^{\prime }\gamma _{1}^{\prime },\beta _{2}^{\prime }\gamma _{2}^{\prime
}}\equiv D_{pp^{\prime }}^{\beta _{1}\gamma _{1},\beta _{2}\gamma
_{2}}U_{p-q,\gamma _{1}\gamma _{1}^{\prime }}U_{p,\beta _{1}\beta
_{1}^{\prime }}^{\ast }U_{p^{\prime },\beta _{2}\beta _{2}^{\prime
}}U_{p^{\prime }-q,\gamma _{2}\gamma _{2}^{\prime }}^{\ast }$ in the
quasi-particle basis using the transformation $U_{k}$. The Fourier transform
of EOM~\eqref{HD} results in%
\begin{equation}
\bar{D}_{kp}^{\beta _{1}\gamma _{1},\beta _{2}\gamma _{2}}(q,\omega _{q})=%
\left[ \frac{1}{[\bar{D}^{(0)}(q,\omega _{q})]^{-1}-\mathcal{V}(q)}\right]
_{k,p}^{\beta _{1}\gamma _{1},\beta _{2}\gamma _{2}},
\end{equation}%
where the bare density-density correlation function reads%
\begin{equation}
\bar{D}_{kp}^{(0)\beta _{1}\gamma _{1},\beta _{2}\gamma _{2}}(q,\omega _{q})=%
\frac{f_{k-q}^{\gamma _{1}}-f_{k}^{\beta _{1}}}{i\omega _{q}-(d_{k,\beta
_{1}}-d_{k-q,\gamma _{1}})}\delta _{\beta _{1}\beta _{2}}\delta _{\gamma
_{1}\gamma _{2}}\delta _{k,p},
\end{equation}%
and the interaction $\mathcal{V}_{k,k^{\prime }}^{\beta _{1}\gamma
_{1},\beta \gamma }(q)=-2V_{k,k^{\prime }}^{\beta _{1}\beta }V_{k^{\prime
}-q,k-q}^{\gamma \gamma _{1}}/N$ of two quasi-particles is determined by $%
V_{k,k^{\prime }}^{\beta _{1}\beta }=(U_{k}^{\dagger }V_{k}^{\dagger
}MV_{k^{\prime }}U_{k^{\prime }})_{\beta _{1}\beta }$. The spectral
decomposition results in%
\begin{equation}
\bar{D}_{p,p^{\prime }}^{\beta _{1}\gamma _{1},\beta _{2}\gamma
_{2}}(q,\omega _{q})=\sum_{\lambda }\chi _{p,\lambda }^{\beta _{1}\gamma
_{1}}(q)\frac{1}{i\omega _{q}-d_{2,\lambda }(q)}\overline{\chi }_{\lambda
,p^{\prime }}^{\beta _{2}\gamma _{2}\ast }(q),
\end{equation}%
where the poles $\pm \left\vert d_{2,\lambda }(q)\right\vert $ of $\bar{D}%
_{p,p^{\prime }}^{\beta _{1}\gamma _{1},\beta _{2}\gamma _{2}}(q,\omega
_{q}) $ appear in pairs due to the particle-hole symmetry of $[\bar{D}%
^{(0)}(q,\omega _{q})]^{-1}-\mathcal{V}(q)$. It turns out that the lowest
band with the dispersion relation $d_{\mathrm{ex}}(q)=\min_{\lambda
}\left\vert d_{2,\lambda }(q)\right\vert $ corresponds to a collective mode,
i.e., the exciton excitation. The bottom of the exciton band is at the $%
\Gamma $-point, and at zero temperature its energy is about $0.08$ for $%
\alpha =0.2$, i.e., one order of magnitude smaller than the bare
quasi-particle gap. {Additionally, the exciton is robust in the presence of the kinetic energy of the bare electron as illustrated in Fig.~\ref{Emf_Eex}.} The wavefunction of the exciton state in the coordinate
space, i.e., the Fourier transformation of $\chi _{p,\lambda }^{\beta
_{1}\gamma _{1}}(q)$, shows that the exciton with the center-of-mass
momentum $q$ is in the bound state of one electron in the conductive band
and one electron in the valence band.

\begin{figure}[htb]
\centering \includegraphics[width=0.8\textwidth]{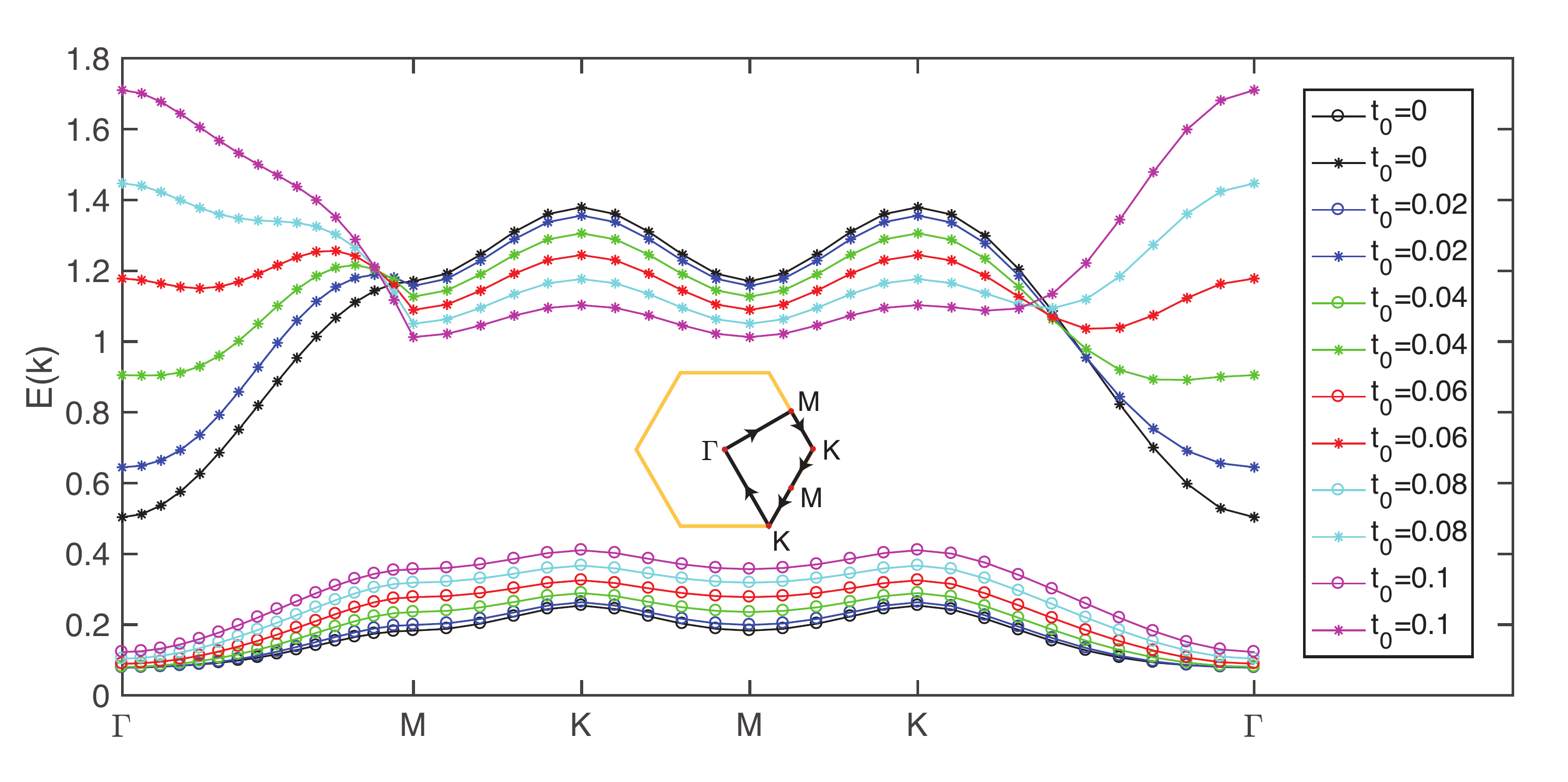}
\caption{{ The bare quasi-particle gap (stars) and the exciton band (circles) obtained via the Hamiltonian $H_{new} = -t_{0} \sum_{\langle i,j\rangle} \left(c_i^{\dagger} c_j +h.c.\right)+ U_{0}\sum_{\varhexagon}\left( Q_{\varhexagon}+\alpha T_{\varhexagon}-1\right)^{2}$ at the effective kinetic strength $t_{0}=0,0.02, 0.04, 0.06,0.08$ and $0.1$, where $\alpha=0.2$, the interaction strength $U_{0}=1$ and the temperature $T=0.04$. The exciton energy is always one order of magnitude smaller than the mean-field gap in the presence of the kinetic energy.}}
\label{Emf_Eex}
\end{figure}

The two-particle correlation function leads to the self-energy correction $%
\Sigma _{k}^{(\text{H})}=U_{k}\bar{\Sigma}_{k}^{(\text{H})}U_{k}^{\dagger }$
via Eq.~\eqref{se}, where $\bar{\Sigma}_{k}^{(\text{H})}$ is the self-energy%
\begin{equation}
\bar{\Sigma}_{k,\alpha \beta }^{(\text{H})}=\frac{4}{N^{2}}\sum_{pp^{\prime
}q}\frac{1-f_{k-q}^{\alpha ^{\prime }}+n_{b}(d_{2,\lambda }(q))}{i\omega
_{k}-d_{k-q,\alpha ^{\prime }}-d_{2,\lambda }(q)}V_{k,p}^{\alpha \beta
_{1}}V_{p-q,k-q}^{\gamma _{1}\alpha ^{\prime }}\chi _{p,\lambda }^{\beta
_{1}\gamma _{1}}(q)\overline{\chi }_{\lambda ,p^{\prime }}^{\beta _{2}\gamma
_{2}\ast }(q)V_{p^{\prime },k}^{\beta _{2}\beta }V_{k-q,p^{\prime
}-q}^{\alpha ^{\prime }\gamma _{2}}
\label{Eq:sigmaH}
\end{equation}%
of the quasi-particle, and $n_{b}(d_{2,\lambda }(q))=1/[\exp (d_{2,\lambda
}(q)/T)-1]$ is the Bose-Einstein distribution.

We get the Matsubara single-particle Green function with the first-order
exciton correction%
\begin{equation}
-\int d\tau e^{i\omega _{k}\tau }\mathcal{h}T_{\tau}c_{k}(\tau )c_{k}^{\dagger }(0)%
\mathcal{i}=\frac{1}{i\omega _{k}-h^{\text{QAH}}(k)-\Sigma _{k}^{(\text{H})}}%
,
\end{equation}%
In the quasi-particle picture, the single-particle Green function becomes%
\begin{equation}
G_{k}(\omega _{k})=-U_{k}^{\dagger }\mathcal{h}c_{k}c_{k}^{\dagger }(\omega
_{k})\mathcal{i}U_{k}=\frac{1}{i\omega _{k}-d_{k}-\bar{\Sigma}_{k}^{(\text{H}%
)}}=\sum_{s}\frac{\bar{Z}_{k,s}}{i\omega _{k}-\bar{\varepsilon}_{k,s}},
\end{equation}%
where $\bar{\varepsilon}_{k,s}$ and $\bar{Z}_{k,s}$ are the poles and
residues of $G_{k}(\omega _{k})$, respectively, which satisfies sum rules
$\sum_{s}\bar{Z}_{k,s}=I_{2}$ and $\sum_{s}\bar{Z}_{k,s}\bar{\varepsilon}%
_{k,s}=d_{k}$.

\begin{figure}[htb]
\centering \includegraphics[width=0.6\textwidth]{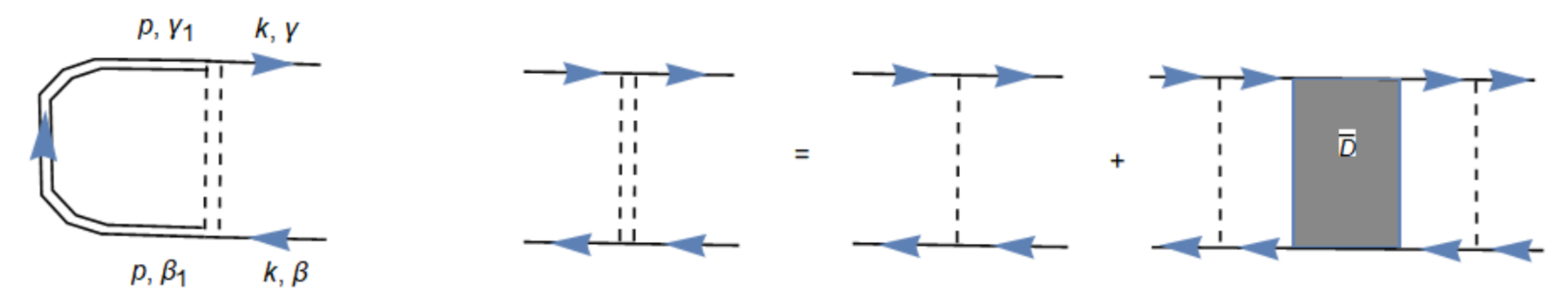}
\caption{The Fock-like self energy $\bar{\Sigma}_{k,\gamma \beta }^{(\text{F})}$.}
\label{Gqah2}
\end{figure}

The Fock correction to the quasi-particle Green function can also be
included as%
\begin{equation}
\bar{\Sigma}_{k,\gamma \beta }^{(\text{F})}=\sum_{p}T_{p,k}^{\beta
_{1}\gamma _{1},\beta \gamma }\left( q=0,\omega _{q}=0\right) T\sum_{\omega
_{p}}\left[ G_{p}^{\gamma _{1}\beta _{1}}\left( \omega _{p}\right) -\bar{G}%
_{p}^{(0)\gamma _{1}\beta _{1}}\left( \omega _{p}\right) \right] ,
\label{Eq:sigmaF}
\end{equation}%
where the interaction $T$-matrix is obtained by%
\begin{equation}
T_{p,k}^{\beta _{1}\gamma _{1},\beta \gamma }\left( q,\omega _{q}\right) =%
\mathcal{V}_{p,k}^{\beta _{1}\gamma _{1},\beta \gamma }+\sum_{k^{\prime
}p^{\prime }}\mathcal{V}_{p,k^{\prime }}^{\beta _{1}\gamma _{1},\mu _{1}\nu
_{1}}\bar{D}_{k^{\prime },p^{\prime }}^{\mu _{1}\nu _{1},\mu \nu }\left(
q,\omega _{q}\right) \mathcal{V}_{p^{\prime },k}^{\mu \nu ,\beta \gamma },
\end{equation}%
and%
\begin{equation}
T\sum_{\omega _{p}}\left[ G_{p}^{\gamma _{1}\beta _{1}}\left( \omega
_{p}\right) -\bar{G}_{p}^{(0)\gamma _{1}\beta _{1}}\left( \omega _{p}\right) %
\right] =\sum_{s}\bar{Z}_{p,s}^{\gamma _{1}\beta _{1}}[f\left( \bar{%
\varepsilon}_{p,s}\right) -1]+\delta _{\gamma _{1}\beta _{1}}\left(
1-f_{k}^{\gamma _{1}}\right)
\end{equation}%
is determined by $G_{k}(\omega _{k})$ and $\bar{G}_{p}^{(0)\gamma _{1}\beta
_{1}}\left( \omega _{p}\right) =\delta _{\gamma _{1}\beta _{1}}/(i\omega
-d_{p,\gamma _{1}})$.

Finally, the full single-particle Green function including the Hartree-Fock-like
corrections becomes%
\begin{equation}
G_{k}^{f}(\omega _{k})=\frac{1}{i\omega _{k}-d_{k}-\bar{\Sigma}_{k}^{(\text{H%
})}-\bar{\Sigma}_{k}^{(\text{F})}}=\sum_{s}\frac{Z_{k,s}}{i\omega
_{k}-\varepsilon _{k,s}},
\end{equation}%
where $\varepsilon _{k,s}$ and $Z_{k,s}$ are the poles and residues of the
spectral function $G_{k}^{f}(\omega _{k})$, respectively. The retarded
Green's function $G_{k}^{(\text{R})}(\omega )=G_{k}^{f}(i\omega
_{k}\rightarrow \omega +i\eta )$ is obtained by the analytic continuation,
which determines the spectral function $A_{k}(\omega )=-$Im$G_{k}^{(\text{R}%
)}(\omega )/\pi $. In our numerical calculation, we choose $\eta =0.004$.

Due to the particle-hole symmetry, $A_{k}^{v}(x)=A_{k}^{c}(-x)$ and we only
focus on the valence. Comparing with the spectral function of the free Green
function $A_{k}^{(0)v}(x)=\delta (x-d_{k}^{v})$, we find that when the
temperature increases, not only the quasi-particle gap is reduced, but the
broadened spectral function $A_{k}^{v}(x)$ even has the non-zero
distribution at the positive frequency, as shown in Fig.~\ref{AwvTQAH}. This is
a strong evidence of the valence band electron dressed by the excitons. The
reduction of the quasi-particle weight and the spectral distribution in the
negative frequency domain indicates the decrease of the current-current
correlation. As shown in Fig.~{\color{blue}3}, with the correction of
the exciton the current-current correlation decreases much faster than the
mean-field result. However, the perturbative expansion fails to predict the
correct critical temperature obtained by XTRG. This is because at the higher
temperature, many excitons are proliferated, which strongly affect the
quasi-particle spectrum as well as the particle-hole spectrum. As a result,
the bare Green function in the calculation of the $T$-matrix should be replaced
by the exact Green function, namely, the self-consistent calculation of the
Green function is required, which will be our future work.
\begin{figure}[htbp]
\center
\subfigure[]{
\includegraphics[width=0.4\textwidth]{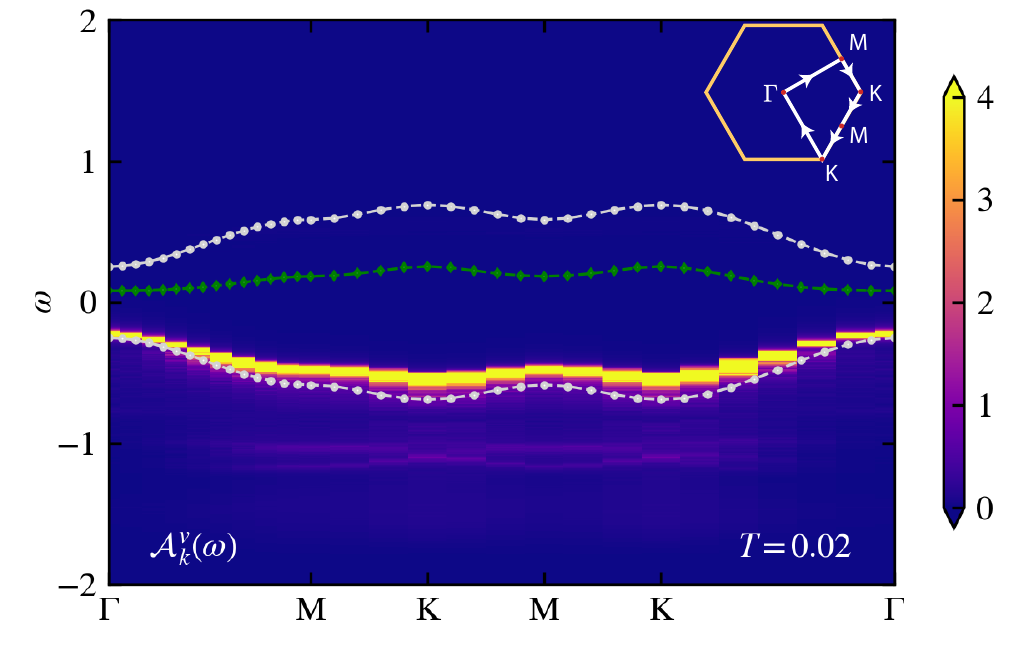}
\label{AwvT002qah} } \quad
\subfigure[]{
\includegraphics[width=0.4\textwidth]{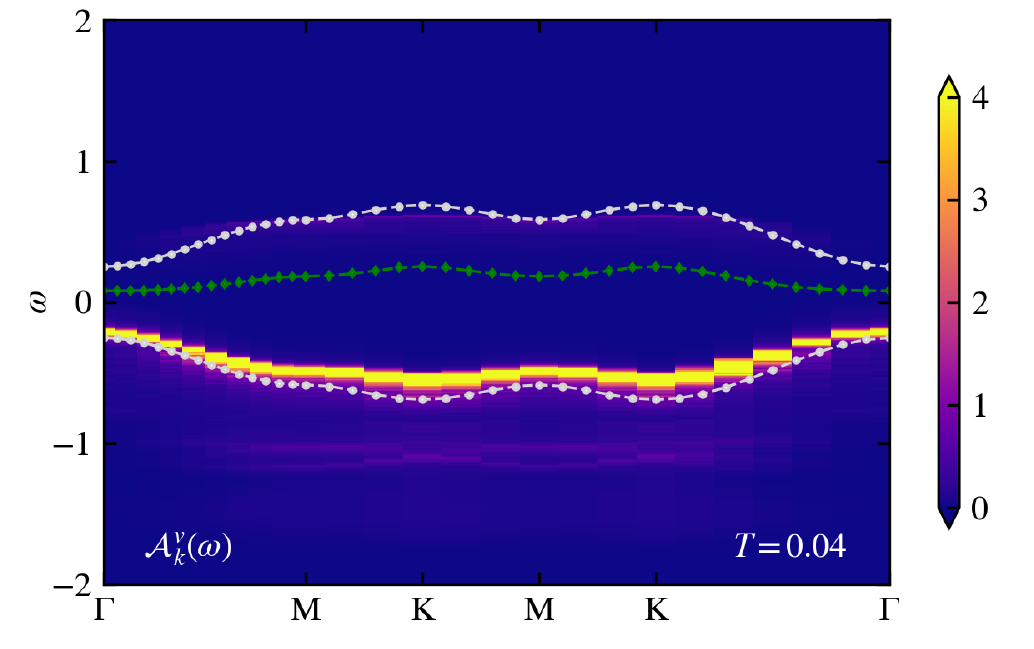}
\label{AwvT004qah} } \quad
\subfigure[]{
\includegraphics[width=0.4\textwidth]{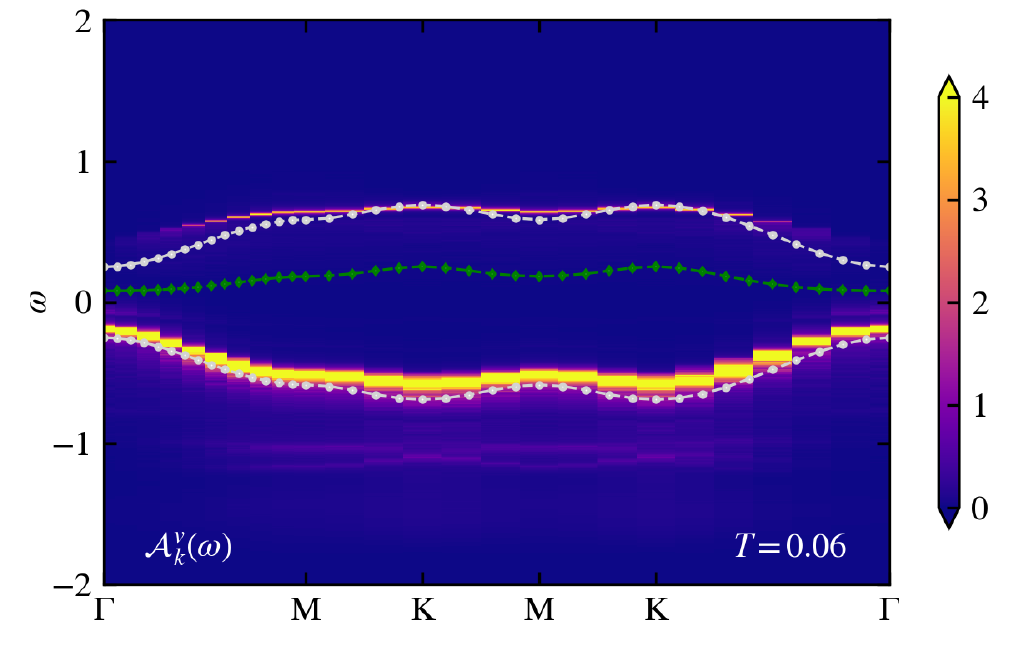}
\label{AwvT006qah} } \quad
\subfigure[]{
\includegraphics[width=0.4\textwidth]{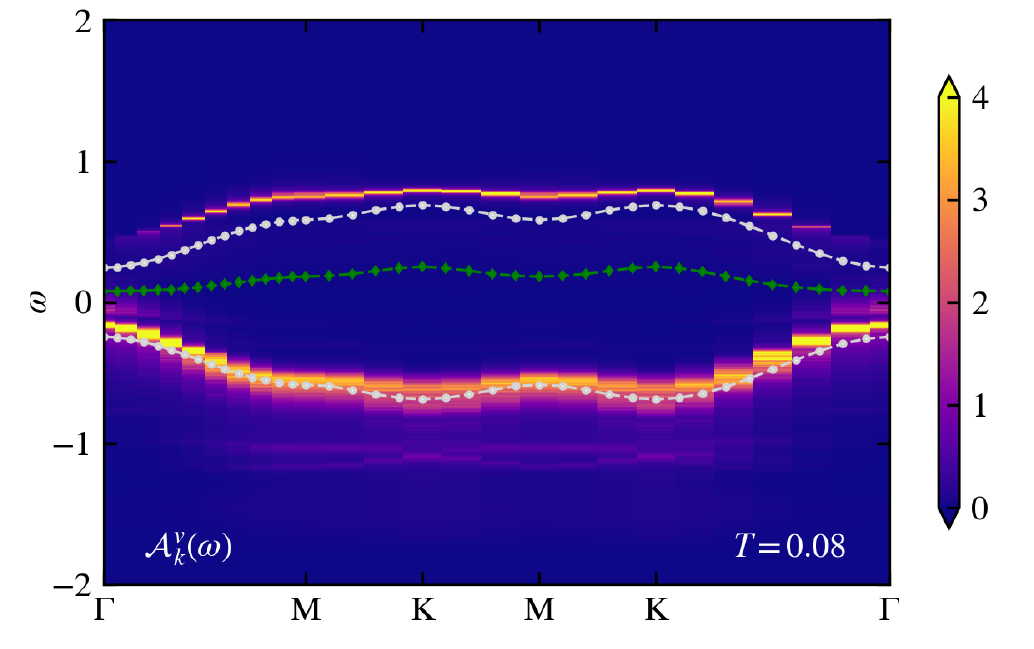}
\label{AwvT008qah} }
\caption{The conductive and valence bands of the free single-particle Green's function (white dots),
the exciton band (green diamonds), and the spectral function $A_{k}^{v}(\omega)$ of the full
single-particle Green's function  on the valence band at $T=0.02, 0.04, 0.06$ and $0.08$ { for $\alpha=0.2$ (QAH phase)}.
As temperature increases, the proliferation of excitons manifests.}
\label{AwvTQAH}
\end{figure}

{
In the CDW phase, the calculation on the $T$-matrix and self-energy corrections can also be
performed in the similar way, which also shows the low lying exciton excitation confirmed by the XTRG results.

The Fourier transformation gives rise to the interacting Hamiltonian
\begin{align}
H_{I} & =\frac{1}{N}\sum_{k_{h},\overline{k}_{h},q_{h}}:c_{k_{h}-q_{h},\alpha'}^{\dagger}\widetilde{V}_{\alpha'\bar{\beta}}^{k_{h}-q_{h},k_{h_{+}}-q_{h};\overline{k}_{h}-q_{h},\overline{k}_{h_{+}}-q_{h}}c_{\overline{k}_{h}-q_{h},\bar{\beta}}c_{\overline{k}_{h},\bar{\alpha}}^{\dagger}\widetilde{V}_{\bar{\alpha}\beta'}^{\overline{k}_{h},\overline{k}_{h_{+}};k_{h},k_{h_{+}}}c_{k_{h},\beta'}:\nonumber \\
 & +\frac{1}{N}\sum_{k_{h},\overline{k}_{h},q_{h}}:c_{k_{h}-q_{h},\alpha'}^{\dagger}\widetilde{V}_{\alpha'\bar{\beta}}^{k_{h}-q_{h},k_{h_{+}}-q_{h};\overline{k}_{h_{+}}-q_{h},\overline{k}_{h}-q_{h}}P_{\bar{\beta}\bar{\beta}_{1}}c_{\overline{k}_{h}-q_{h},\bar{\beta}_{1}}c_{\overline{k}_{h},\bar{\alpha}_{1}}^{\dagger}P_{\bar{\alpha}_{1}\bar{\alpha}}\widetilde{V}_{\bar{\alpha}\beta'}^{\overline{k}_{h_{+}},\overline{k}_{h};k_{h},k_{h_{+}}}c_{k_{h},\beta'}:
\end{align}
 in the momentum space $
c_{k_{h}}=(a_{k_{h}},b_{k_{h}},a_{k_{h_{+}}},b_{k_{h_{+}}})$, where the $4\times4$ interacting matrix $\widetilde{V}^{k,k';p,p'}=\left(\begin{array}{cc}
V_{k}^{\dagger}MV_{p} & 0\\
0 & V_{k'}^{\dagger}MV_{p'}
\end{array}\right)$ and the permutation matrix $P=\sigma^{x}\otimes I_{2}$.

The two-particle spectrum is also obtained by the ladder diagram approximation. The self-energy correction, e.g., the Hartree term, becomes
\begin{align}
\Sigma_{k_{h},\alpha\beta}^\text{{(H)CDW}} & =-\left(\frac{2}{N}\right)^{2}\sum_{p_{h}p_{h}'q_{h}}\frac{1}{\beta}\sum_{\omega_{q_{h}}}G_{k_{h}-q_{h},\alpha_{1}\alpha_{2}}^\text{{(0)CDW}}(\omega_{k_{h}}-\omega_{q_{h}})D_{p_{h}p_{h}^{'}}^{\beta_{1}\gamma_{1},\beta_{2}\gamma_{2}}(q_{h},\omega_{q_{h}})\\
 & [\widetilde{V}_{\alpha\beta_{1}}^{k_{h},k_{h_{+}};p_{h}^{'},p_{h_{+}}^{'}}\widetilde{V}_{\gamma_{1}\alpha_{1}}^{p_{h}^{'}-q_{h},p_{h_{+}}^{'}-q_{h};k_{h}-q_{h},k_{h_{+}}-q_{h}}+P_{\alpha\alpha'}\widetilde{V}_{\alpha'\beta_{1}}^{k_{h_{+}},k_{h};p_{h}^{'},p_{h_{+}}^{'}}\widetilde{V}_{\gamma_{1}\alpha_{1}^{'}}^{p_{h}^{'}-q_{h},p_{h_{+}}^{'}-q_{h};k_{h_{+}}-q_{h},k_{h}-q_{h}}P_{\alpha_{1}^{'}\alpha_{1}}]\nonumber \\
 & [\widetilde{V}_{\beta_{2}\beta}^{p_{h},p_{h_{+}};k_{h},k_{h_{+}}}\widetilde{V}_{\alpha_{2}\gamma_{2}}^{k_{h}-q_{h},k_{h_{+}}-q_{h};p_{h}-q_{h},p_{h_{+}}-q_{h}}+P_{\beta_{2}\beta_{2}^{'}}\widetilde{V}_{\beta_{2}^{'}\beta}^{p_{h_{+}},p_{h};k_{h},k_{h_{+}}}\widetilde{V}_{\alpha_{2}\gamma_{2}^{'}}^{k_{h}-q_{h},k_{h_{+}}-q_{h};p_{h_{+}}-q_{h},p_{h}-q_{h}}P_{\gamma_{2}^{'}\gamma_{2}}],\nonumber
\end{align}
where the free single-particle Green function $G_{k_{h}}^{(0)\text{CDW}}(\omega_{k_{h}})=1/(i\omega_{k_h}-h^{\text{CDW}}(k_{h}))$.
The exciton spectrum can be determined by the density-density correlation function
\begin{align}
D_{p_{h}p_{h}^{'}}^{\beta_{1}\gamma_{1},\beta_{2}\gamma_{2}}(q_{h},\omega_{q_{h}}) & =\int_{0}^{\beta}d\tau D_{p_{h}p_{h}^{'}}^{\beta_{1}\gamma_{1},\beta_{2}\gamma_{2}}(q_{h},\tau)e^{i \omega_{q_{h}} \tau},\\
D_{p_{h}p_{h}^{'}}^{\beta_{1}\gamma_{1},\beta_{2}\gamma_{2}}(q_{h},\tau) & =-\mathcal{h}T_{\tau}c_{p_{h}-q_{h},\gamma_{1}}^{\dagger}(\tau)c_{p_{h},\beta_{1}}(\tau)c_{p_{h}^{'},\beta_{2}}^{\dagger}(0)c_{p_{h}^{'}-q_{h},\gamma_{2}}(0)\mathcal{i}.
\end{align}

It follows from the Heisenberg EOM that
\begin{align}
-\partial_{\tau}D_{\bar{p}_{h}p_{h}}^{\beta_{1}\gamma_{1},\beta_{2}\gamma_{2}}(q_{h},\tau) & =(\mathcal{h}c_{p_{h}-q_{h},\gamma_{1}}^{\dagger}c_{p_{h}-q_{h},\gamma_{2}}\mathcal{i}_{G}\delta_{\beta_{1},\beta_{2}}-\mathcal{h}c_{p_{h},\beta_{2}}^{\dagger}c_{p_{h},\beta_{1}}\mathcal{i}_{G}\delta_{\gamma_{1},\gamma_{2}})\delta_{\bar{p}_{h},p_{h}}\delta(\tau)+M_{\bar{p}_{h},\bar{k}_{h}}^{\beta_{1}\gamma_{1},\beta\gamma}(q_{h})D_{\bar{k}_{h}p_{h}}^{\beta\gamma,\beta_{2}\gamma_{2}}(q_{h},\tau),\label{EOMcdw}
\end{align}
where
\begin{align}
 & M_{\bar{p}_{h},\bar{k}_{h}}^{\beta_{1}\gamma_{1},\beta\gamma}(q_{h})=(h_{\beta_{1}\beta}^{\text{CDW}}(\bar{p}_{h})\delta_{\gamma_{1}\gamma}-\delta_{\beta_{1}\beta}h_{\gamma\gamma_{1}}^{\text{CDW}}(\bar{p}_{h}-q_{h}))\delta_{\bar{p}_{h},\bar{k}_{h}}\\
 & +\frac{2}{N}[\left(\widetilde{V}_{\gamma\gamma_{1}}^{\bar{k}_{h}-q_{h},\bar{k}_{h_{+}}-q_{h};\bar{p}_{h}-q_{h},\bar{p}_{h_{+}}-q_{h}}\widetilde{V}_{\bar{\alpha}\beta}^{\bar{p}_{h},\bar{p}_{h_{+}};\bar{k}_{h},\bar{k}_{h_{+}}}+\widetilde{V}_{\gamma\bar{\beta}_{1}}^{\bar{k}_{h}-q_{h},\bar{k}_{h_{+}}-q_{h};\bar{p}_{h_{+}}-q_{h},\bar{p}_{h}-q_{h}}P_{\bar{\beta}_{1}\gamma_{1}}P_{\bar{\alpha}\bar{\alpha}_{1}}\widetilde{V}_{\bar{\alpha}_{1}\beta}^{\bar{p}_{h_{+}},\bar{p}_{h};\bar{k}_{h},\bar{k}_{h_{+}}}\right)\mathcal{h}c_{\bar{p}_{h},\bar{\alpha}}^{\dagger}c_{\bar{p}_{h},\beta_{1}}\mathcal{i}_{G}\nonumber \\
 & -\left(\widetilde{V}_{\gamma\bar{\beta}}^{\bar{k}_{h}-q_{h},\bar{k}_{h_{+}}-q_{h};\bar{p}_{h}-q_{h},\bar{p}_{h_{+}}-q_{h}}\widetilde{V}_{\beta_{1}\beta}^{\bar{p}_{h},\bar{p}_{h_{+}};\bar{k}_{h},\bar{k}_{h_{+}}}+\widetilde{V}_{\gamma\bar{\beta}_{1}}^{\bar{k}_{h}-q_{h},\bar{k}_{h_{+}}-q_{h};\bar{p}_{h_{+}}-q_{h},\bar{p}_{h}-q_{h}}P_{\bar{\beta}_{1}\bar{\beta}}P_{\beta_{1}\bar{\alpha}_{1}}\widetilde{V}_{\beta_{1}\beta}^{\bar{p}_{h_{+}},\bar{p}_{h};\bar{k}_{h},\bar{k}_{h_{+}}}\right)\mathcal{h}c_{\bar{p}_{h}-q_{h},\gamma_{1}}^{\dagger}c_{\bar{p}_{h}-q_{h},\bar{\beta}}\mathcal{i}_{G}].\nonumber
\end{align}

We define the correlation function $\bar{D}_{p_{h}'p_{h}}^{\beta_{1}\gamma_{1},\beta_{2}\gamma_{2}}(q,\tau)\equiv D_{p_{h}'p_{h}}^{\beta_{1}'\gamma_{1}',\beta_{2}'\gamma_{2}'}(q,\tau)U_{p_{h}'-q,\gamma_{1}'\gamma_{1}}U_{p_{h}',\beta_{1}\beta_{1}'}^{\dagger}U_{p_{h},\beta_{2}'\beta_{2}}U_{p_{h}-q,\gamma_{2}\gamma_{2}'}^{\dagger}$ in the quasi-particle basis via the transformation $U_{p_h}$ in the CDW phase. The Fourier transform of EOM~\eqref{EOMcdw} results in
\begin{align}
\bar{D}_{p_{h}'p_{h}}^{\beta_{1}\gamma_{1},\beta_{2}\gamma_{2}}\left(q_{h},\omega_{q_{h}}\right) & =\left(\frac{1}{i\omega_{q_{h}}-\bar{M}(q_{h})}\right)_{\bar{p}_{h}k_{h}}^{\beta_{1}\gamma_{1},\beta\gamma}\bar{N}{}_{k_{h}p_{h}}^{\beta\gamma,\beta_{2}\gamma_{2}}\left(q_{h},\omega_{q_{h}}\right),\\
\bar{M}_{\bar{p}_{h},\bar{k}_{h}}^{\beta_{1}\gamma_{1},\beta\gamma}\left(q_{h}\right) & =\left(d_{\bar{p}_{h},\beta_{1}\beta}\delta_{\gamma_{1},\gamma}-\delta_{\beta_{1},\beta}d_{\bar{p}_{h}-q_{h},\gamma_{1}\gamma}\right)\delta_{\bar{p}_{h},\bar{k}_{h}}+\mathcal{V}_{\bar{p}_{h},\bar{k}_{h}}^{\left(h,h_{+}\right)\beta_{1}\gamma_{1},\beta\gamma}\left(q_{h}\right)\left(f_{\bar{p}_{h}-q_{h}}^{\gamma_{1}}-f_{\bar{p}_{h}}^{\beta_{1}}\right),\\
\bar{N}_{\bar{p}_{h}p_{h}}^{\beta_{1}\gamma_{1},\beta_{2}\gamma_{2}}\left(q,\omega_{q}\right) & =\left(f_{p_{h}-q}^{\gamma_{1}}-f_{p_{h}}^{\beta_{1}}\right)\delta_{p_{h},\bar{p}_{h}}\delta_{\beta_{1},\beta_{2}}\delta_{\gamma_{1},\gamma_{2}},
\end{align}
where the interaction
\begin{align}
\mathcal{V}_{\bar{p}_{h},\bar{k}_{h}}^{\left(h,h_{+}\right)\beta_{1}\gamma_{1},\beta\gamma}\left(q_{h}\right)=\frac{2}{N}\left(-V_{\bar{k}_{h}-q_{h},\bar{p}_{h}-q_{h}}^{\left(h,h_{+}\right)\gamma\gamma_{1}}V_{\bar{p}_{h},\bar{k}_{h}}^{\left(h,h_{+}\right)\beta_{1}\beta}-V_{\bar{k}_{h}-q_{h},(P)\bar{p}_{h}-q_{h}}^{\left(h,h_{+}\right)\gamma\gamma_{1}}V_{(P)\bar{p}_{h},\bar{k}_{h}}^{\left(h,h_{+}\right)\beta_{1}\beta}\right) \end{align}
of two quasi-particles is determined by
$V_{\bar{k}_{h}-q_{h},\bar{k}_{h}}^{\left(h,h_{+}\right)\gamma\beta}\equiv U_{\bar{k}_{h}-q_{h},\gamma\gamma'}^{\dagger}\widetilde{V}_{\gamma'\beta'}^{\bar{k}_{h}-q_{h},\bar{k}_{h_{+}}-q_{h};\bar{k}_{h},\bar{k}_{h_{+}}}U_{\bar{k}_{h},\beta'\beta}$ and $V_{(P)\bar{p}_{h},\bar{k}_{h}}^{\left(h,h_{+}\right)\beta_{1}\beta}\equiv U_{\bar{p}_{h},\beta_{1}\gamma'}^{\dagger}P_{\gamma'\alpha_{1}}\\ \widetilde{V}_{\alpha_{1}\beta'}^{\bar{p}_{h_{+}},\bar{p}_{h};\bar{k}_{h},\bar{k}_{h_{+}}}U_{\bar{k}_{h},\beta'\beta}$.
The spectral decomposition results in
\begin{equation}
\bar{D}_{p_{h}'p_{h}}^{\beta_{1}\gamma_{1},\beta_{2}\gamma_{2}}\left(q_{h},\omega_{q_{h}}\right)=\chi_{\bar{p}_{h},\lambda}^{\beta_{1}\gamma_{1}}\left(q_{h}\right)\frac{1}{i\omega_{q_{h}}-d_{2,\lambda}\left(q_{h}\right)}\bar{\chi}_{p_{h},\lambda}^{\beta\gamma*}\left(q_{h}\right).
\end{equation}

In the CDW phase, the lowest
band with the dispersion relation $d_{\mathrm{ex}}(q)=\min_{\lambda
}\left\vert d_{2,\lambda }(q)\right\vert $ also corresponds to a collective mode,
i.e., the exciton excitation. The lowest energy of the exciton is about 0.15 for $\alpha=0.1$, which is smaller then the quasi-particle gap, similar to what happens in the QAH phase. However, the exciton band structure in the CDW phase is significantly distinct from that in the QAH phase due to the different spatial symmetries, as shown in Fig.~\ref{Fig:EgapfullK}.

\begin{figure}[htb]
\includegraphics[width=.5\linewidth]{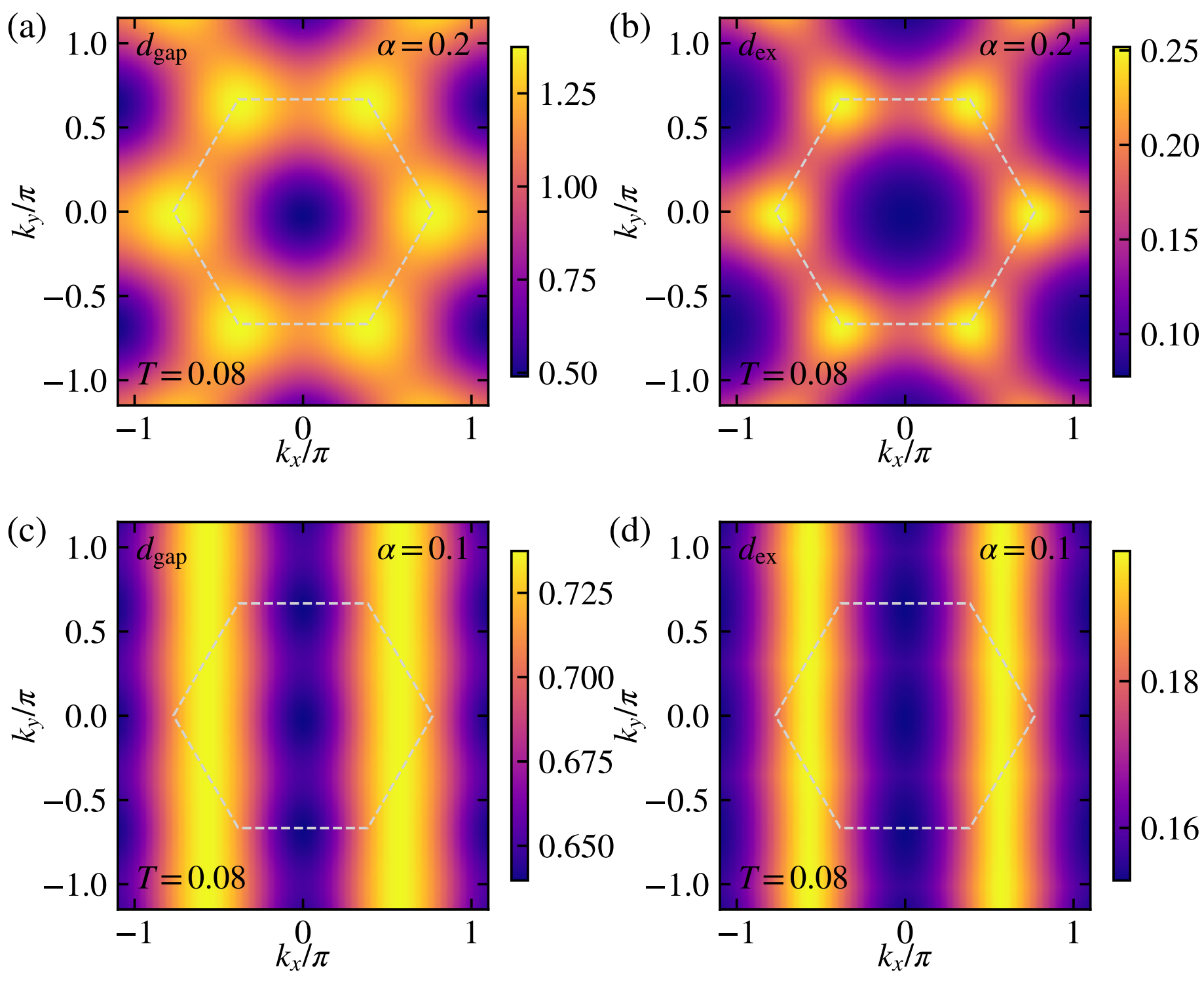}
\caption{{The quasi-particle gap $d_{gap}=2 d_{k,c}( 2 E_{k})$ and the exciton band $d_{ex}$ in the QAH (CDW) phase, where $\alpha =0.2 (0.1) $. The exciton energy is one order of magnitude smaller than the
quasi-particle gap. Besides, the bands in the CDW phase have the reduce period in the $y$ direction.}}
\label{Fig:EgapfullK}
\end{figure}

The self-energy corrections, i.e., the Hartree-Fock corrections $\bar{\Sigma}_{k}^{\text{(H)CDW}}$ and $\bar{\Sigma}_{k}^{\text{(F)CDW}}$, can also be obtained in the similar way as Eq.~\eqref{Eq:sigmaH} and Eq.~\eqref{Eq:sigmaF}. One only has to change the corresponding quasi-particle Green function and the interaction matrix to those in the CDW phase, which results in the full single-particle Green function
\begin{equation}
G^{\text{CDW}}_{k_h}\left(\omega_{k_h}\right)=-\mathcal{h}c_{k_{h}}c_{k_{h}}^{\dagger}(\omega_{k_{h}})\mathcal{i}=\frac{1}{i\omega_{k_{h}}-U_{k_h}^{\dagger}h^{\text{CDW}}(k_{h})U_{k_h}-\bar{\Sigma}_{k_{h}}^{\text{(H)CDW}}-\bar{\Sigma}_{k_{h}}^{\text{(F)CDW}}}
\end{equation}
and the spectral functions $A_{k}^{\beta}(\omega)$. Due to the symmetries mentioned in the last section, there are two degenerate valence bands and two degenerate conductance bands, as a result, four single-particle spectral functions are obtained in the CDW phase. Due to the particle-hole symmetry, we only focus on the spectral function $A_{k}^{v}(\omega)$ for one of the degenerate valence bands.
When the temperature increases, the quasi-particle gap is reduced and the exciton mode assists the valence electron to tunnel across the band gap, as shown in Fig.~\ref{AwvTCDW}.

Via choosing the specific path in the momentum space, as illustrated in Fig.~\ref{Fig:Akw}, the difference of the spectral function between the QAH and CDW phases are explicitly displayed. In our case, the period of the spectral function in the CDW phase is reduced by half in the $y$ direction. In the QAH phase, the translational symmetry $P_{\overrightarrow{a}_{i}}$ ($i=1,2,3$) are preserved. However, the stripe in the CDW phase spontaneously breaks the the translational symmetry $P_{\overrightarrow{a}_{i}}$ ($i=1,2,3$) and preserves the translational symmetry $P_{2 \overrightarrow{a}_{i}}$ ($i=1,2,3$), as a result, the period in the momentum space is reduced.

\begin{figure}[htbp]
\center
\subfigure[]{
\includegraphics[width=0.4\textwidth]{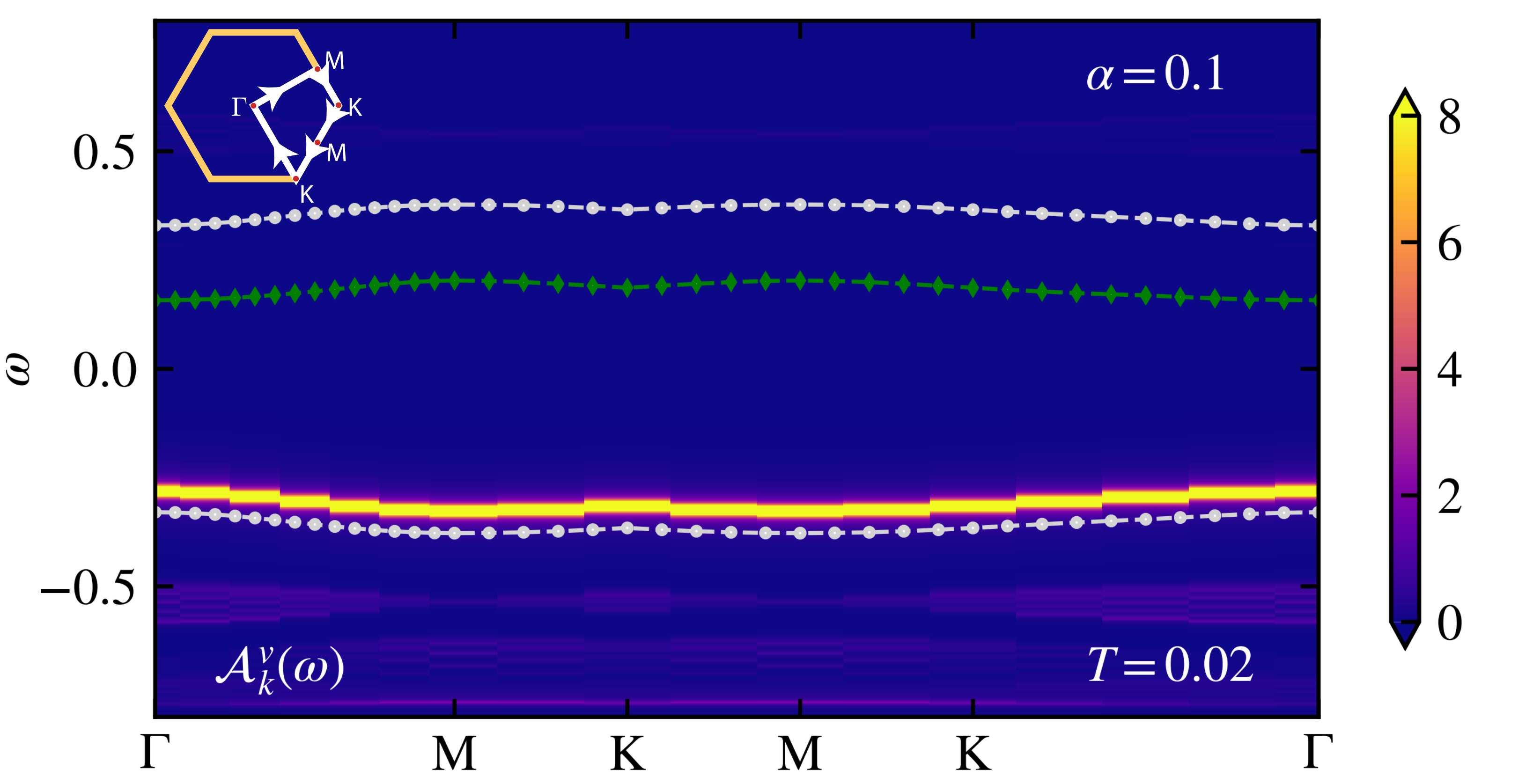}
\label{AwvT002cdw} } \quad
\subfigure[]{
\includegraphics[width=0.4\textwidth]{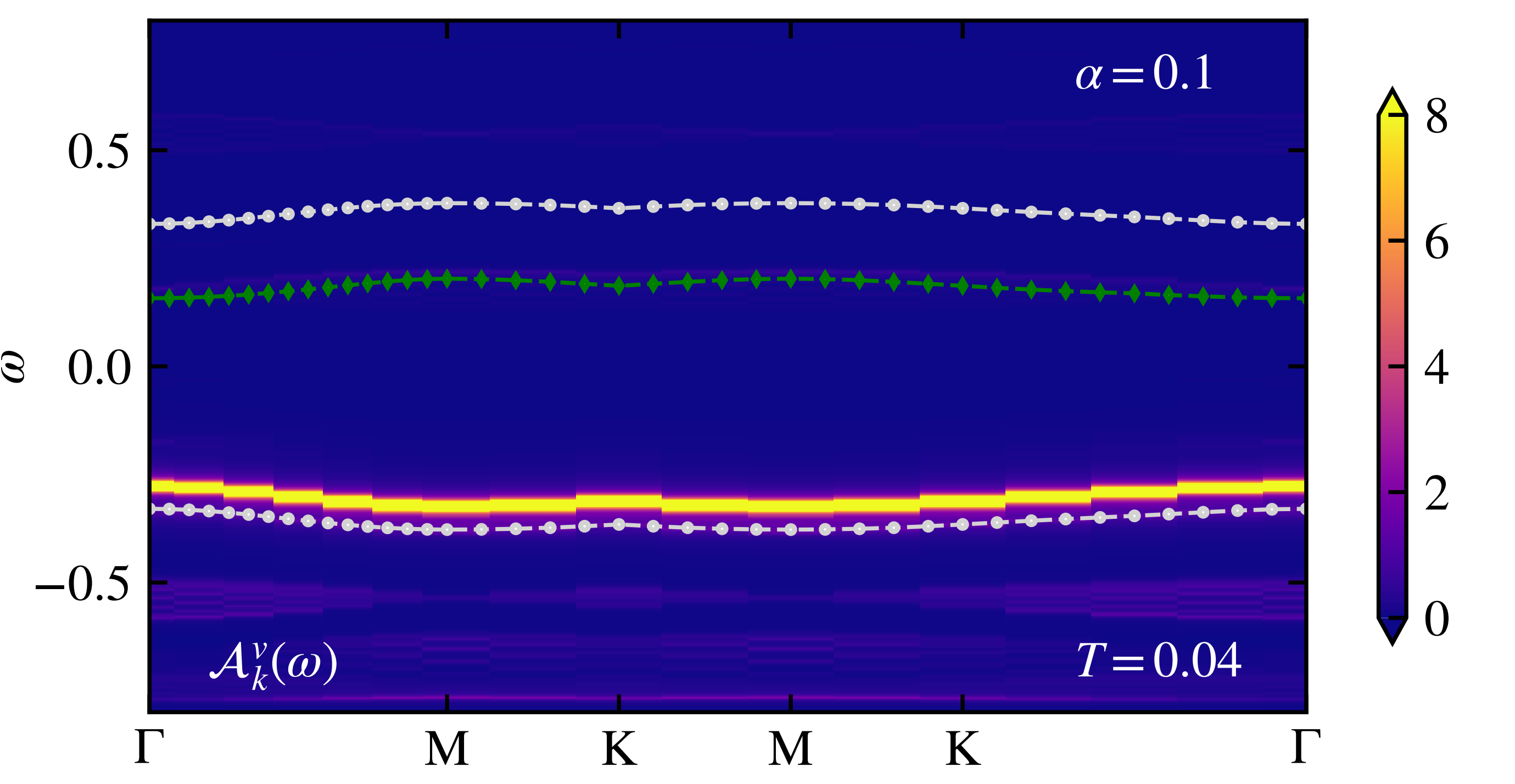}
\label{AwvT004cdw} } \quad
\subfigure[]{
\includegraphics[width=0.4\textwidth]{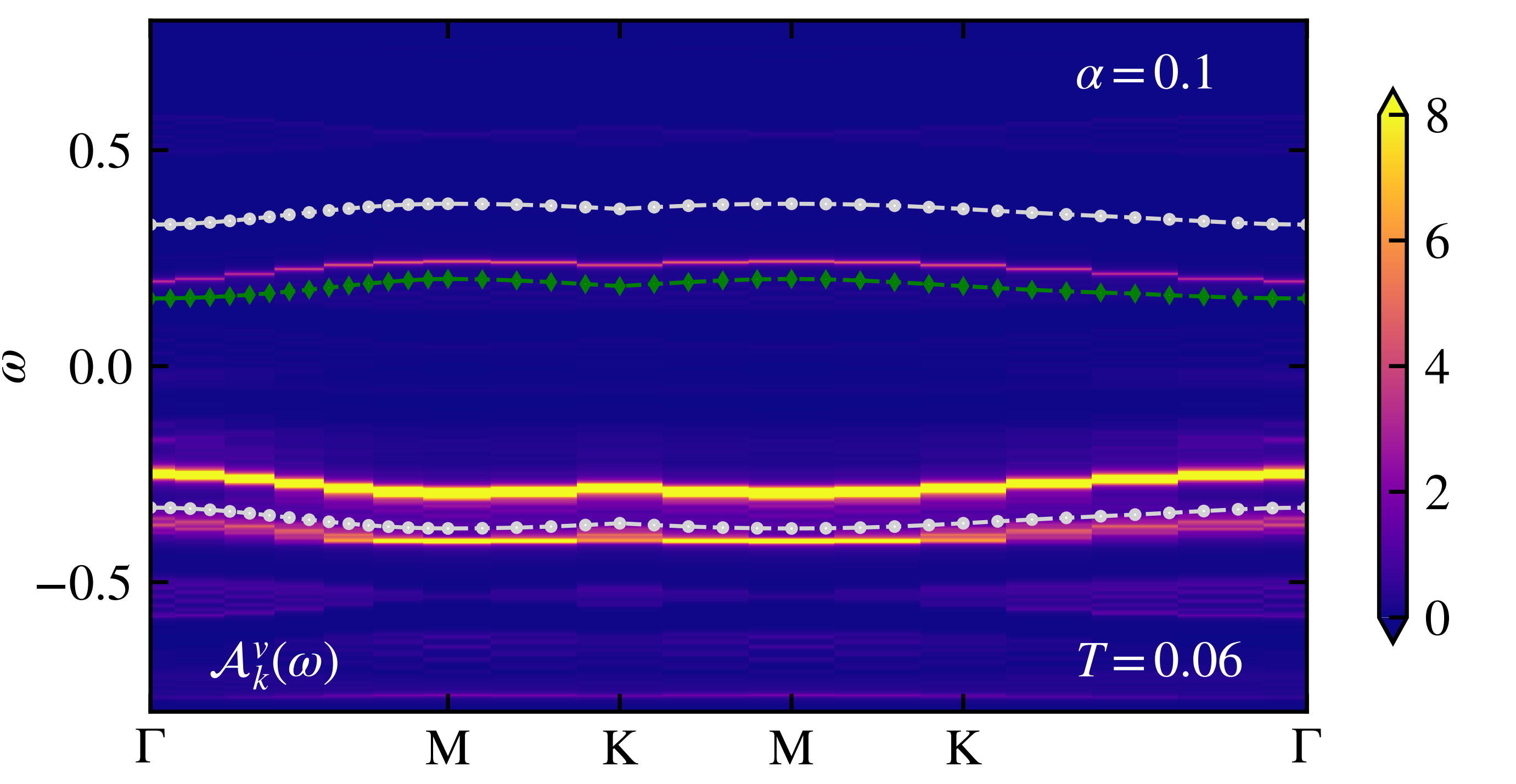}
\label{AwvT006cdw} } \quad
\subfigure[]{
\includegraphics[width=0.4\textwidth]{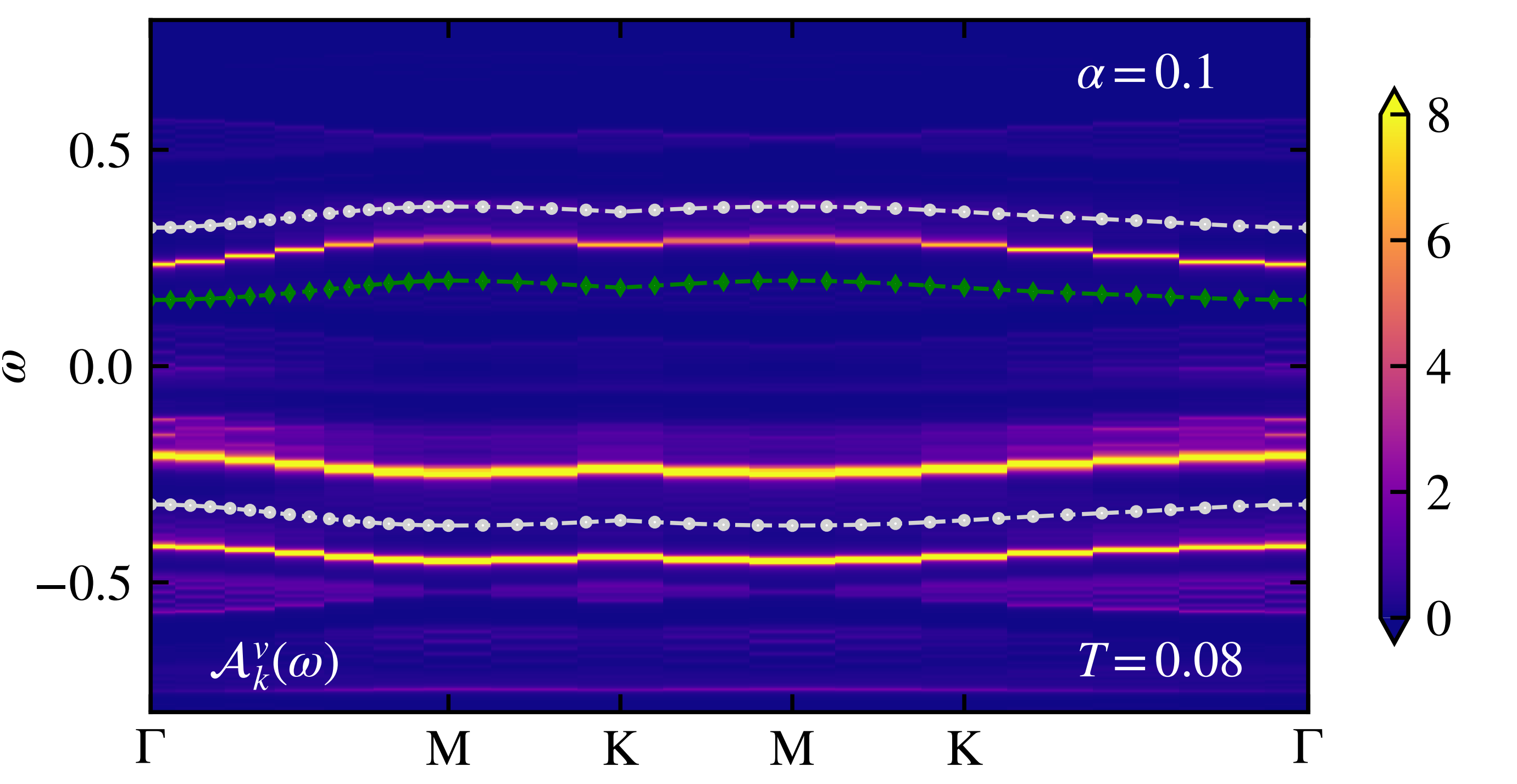}
\label{AwvT008cdw} }
\caption{{The conductive and valence bands (white dots), the exciton band (green diamonds), and the spectral function $A_{k}^{v}(\omega)$ at $T=0.02, 0.04, 0.06$ and $0.08$ for $\alpha=0.1$ in the CDW  phase.}}
\label{AwvTCDW}
\end{figure}
\begin{figure}[htb]
\centering \includegraphics[width=0.82\textwidth]{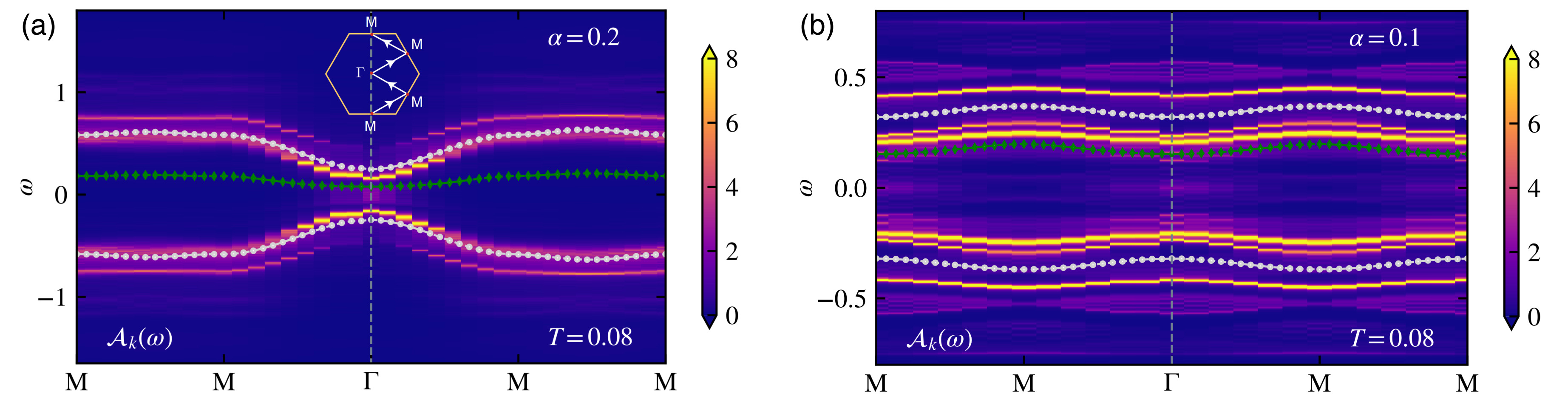}
\caption{{ The single-particle spectral function $\mathcal{A}_k(\omega) = \mathcal{A}^v_k(\omega) +\mathcal{A}^c_k(\omega) $ for both the valence and conductive bands.
(a) $\alpha =0.2 $ (QAH phase) and (b) $\alpha =0.1$ (CDW phase). The period of the spectral function in the CDW phase is reduced by half.}}
\label{Fig:Akw}
\end{figure}
}

\subsection{Section IX: Numerical results on Haldane-Hubbard model}

\begin{figure}[!t]
\includegraphics[angle=0,width=\linewidth]{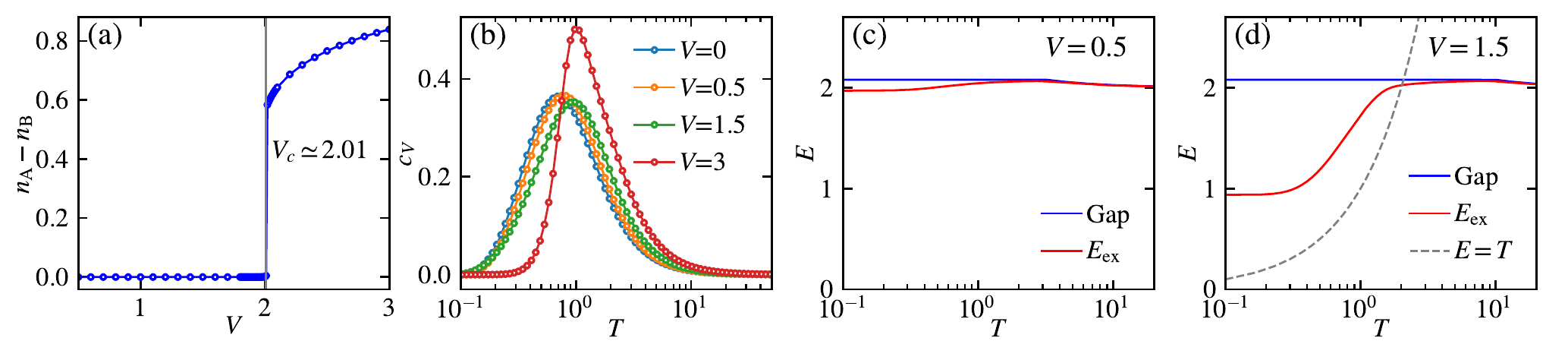}
\caption{In a YC4$\times$12$\times$2 Haldane-Hubbard model, (a) sub-lattice charge density difference $n_{\mathrm{A}} - n_{\mathrm{B}}$ obtained from DMRG calculation are shown versus nearest neighboring Coulomb repulsion strength $V$. (b) Specific heat $c_V$ obtained from XTRG calculation is shown versus temperature $T$ for different cases, i.e. $V=0, 0.5, 1.5$ (CI) and $V=3$ (CDW). 
(c,d) Band gap and exciton energy obtained from field-theoretical calculation are shown versus temperature $T$ for $V=0.5, 1.5$.  }
\label{Fig:FigS20}
\end{figure}

{
We perform DMRG ($T=0$) and XTRG ($T>0$) calculations of the 
spinless Haldane-Hubbard model (HHM) whose Hamiltonian reads
\begin{equation}
H = -t\sum_{\langle ij\rangle} ( c^\dag_i c^{\,}_j + \mathrm{h.c.} ) 
-t'\sum_{\langle\langle ij\rangle\rangle} (e^{\mathrm{i}\phi_{ij}} 
c^\dag_i c^{\,}_j +  \mathrm{h.c.} ) + V\sum_{\langle ij\rangle} 
(n_i-\frac{1}{2})(n_j-\frac{1}{2}),
\label{Eq:HHM}
\end{equation}
where $t'=0.2t, \phi_{ij}=\frac{\pi}{2}$ and the direction of 
next-nearest-neighbor (NNN) pair $\langle\langle ij\rangle\rangle$ 
follows the standard Haldane model~\cite{Haldane1988}, 
and introduce the nearest-neighbor (NN) repulsion as in, 
e.g., Ref.~\cite{CanShao2021}. The model in Eq.~(\ref{Eq:HHM})
keeps the particle-hole symmetry and thus guarantees 
half filling in the finite-temperature XTRG calculations below. 
$t=1$ is set as the energy scale below.

In \Fig{Fig:FigS20}(a), we show the DMRG results of charge density difference 
$(n_{\mathrm{A}}-n_{\mathrm{B}})\equiv \frac{2}{N}(\sum_{i\in \mathrm{A}} 
n_i - \sum_{i\in \mathrm{B}} n_i)$ between sublattice A and B, 
which serves as an order parameter of the charge density wave 
(CDW) phase with large $V$. Around $V_c\simeq2.01$, we find 
a sudden jump from zero to finite value in $n_{\mathrm{A}}
-n_{\mathrm{B}}$, which suggests a first-order phase 
transition there. 
Note that, this transition $V_c$ is half of the corresponding value in 
Ref.~\cite{CanShao2021}, due to the fact that, for the spinless fermion here, 
half of the Coulomb interaction terms (i.e. those $n_{i\uparrow}n_{j\downarrow}$ terms) are discarded.
  
In Fig.~\ref{Fig:FigS20}(b) we present the XTRG results of the specific
heat in the cases with $V=0, 0.5, 1.5$ (CI) and $V=3$ (CDW).
As temperature rises up, in both phases, i.e., CDW and CI, the specific
heat establishes round peaks (instead of divergent ones). 
Note that, with the increasing interaction strength from $V=0$ to $V=1.5$ 
in the CI phase, the position of the specific heat round peaks moves to 
higher temperatures.
It contradicts the exciton proliferation picture, where the larger binding energy induced by interactions (the extremely lower exciton excitation mode) gives rise to the lower position of the specific round peaks as $V$ increases.


As shown in Fig.~\ref{Fig:FigS20}(c-d), 
the fluctuation spectra analysis confirms that the exciton
bounded states are hardly occupied in the CI phase of the HHM, in stark
contrast to the TBG model where exciton proliferation occurs in the QAH
phase. The reason why excitons are rarely occupied in the HHM is that
the exciton energy scale is always higher than the system temperature.
Here are two cases: (a) When the nearest-neighbor interaction $V$ is
relatively small (say, $V=0.5$), as illustrated in \Fig{Fig:FigS20}(c), 
the Haldane term predominates and the exciton energy remains roughly 
at the same value of the energy gap $2$ even if the temperature increases.
(b) It is even more interesting for a larger $V$ (say, $V=1.5$).
As temperature increases, the attractive interaction between electrons and holes is screened 
by the individual thermal excitations, resulting in a reduced exciton binding energy 
and a larger spatial distribution of the exciton wavefunction.
For instance, the exciton energies are $E_{ex} \simeq
0.8,1.1$, and $1.7$ for $V=1.5$ at temperatures $T=0,0.5$ and $1$, respectively.
As shown in \Fig{Fig:FigS20}(d), the exciton energy is always higher than the corresponding temperature, 
therefore the exciton modes are hardly populated by thermal fluctuations. 
This explains why the specific heat
curves in \Fig{Fig:FigS20}(b) change only slightly as $V$ increases from 0 to 1.5.

With the above numerical calculations and field-theoretical analysis on HHM,
we conclude a fundamental difference in terms of distinct exciton energies
and thermal properties when compared to the flat-band twisted bilayer
graphene systems. The uniqueness of the latter originates from the
interaction-driven, emergent, single-particle band in the strongly coupling limit.
}

\end{widetext}
\end{document}